\documentclass[3p,times,authoryear]{elsarticle}
\usepackage{latexsym}

\usepackage{indentfirst}
\usepackage[english]{babel}
\usepackage{cancel}

\usepackage{epstopdf}
\usepackage{lipsum}
\usepackage{array}
\usepackage{color} 
\usepackage{tabularx}
\usepackage{graphicx} 
\usepackage{amsmath}
\usepackage{amssymb}
\usepackage{amsfonts}	
\usepackage{moreverb}
\usepackage{dsfont}
\usepackage{tipa}
\usepackage{upgreek}
\usepackage{bm}
\usepackage{multirow}
\usepackage{soul}
\usepackage{ textcomp }
\usepackage[dvipsnames]{xcolor}
\usepackage{mathtools}
\usepackage{setspace}
\usepackage{algorithm}
\usepackage{algorithmic}
\usepackage{caption}
\usepackage{subcaption}
\usepackage{booktabs}
\usepackage{dsfont}
\allowdisplaybreaks

\usepackage{pgfplots}
\pgfplotsset{compat=1.15}
\usetikzlibrary{arrows}

\definecolor{ffqqtt}{rgb}{1.,0.,0.2}
\definecolor{ffqqqq}{rgb}{1.,0.,0.}

\newcommand{\figA}{	
\begin{tikzpicture}[scale=0.3, line cap=round,line join=round,>=triangle 45,x=1.0cm,y=1.0cm]
	\clip(-7.9,-7.9) rectangle (15.39,9.22);
	\draw [line width=2.pt] (0.,0.) circle (7.211102550927979cm);
	\draw [line width=2.pt] (0.,0.)-- (4.,6.);
	\draw [line width=2.pt] (0.,0.)-- (7.18050764726596,-0.6635585336314104);
	\draw [shift={(0.,0.)},line width=5.2pt,dash pattern=on 2pt off 2pt,color=ffqqtt]  plot[domain=-0.09214937134323709:0.982793723247329,variable=\t]({1.*7.211102550927979*cos(\t r)+0.*7.211102550927979*sin(\t r)},{0.*7.211102550927979*cos(\t r)+1.*7.211102550927979*sin(\t r)});
	\draw [line width=2.pt,dash pattern=on 2pt off 2pt] (0.,0.) circle (5.596924979912561cm);
	\draw (0.5019648792151653,5.690286979965797) node[anchor=north west] {$R_0$};
	\draw (3.944971549095065,8.435386892437611) node[anchor=north west] {$R_0+\eta$};
	\draw (0.4554377620546261,-0.5443467195464587) node[anchor=north west] {$\sqrt{\gamma}(R_0)$};
	\draw (6.969234164530112,4.154892113668002) node[anchor=north west] {$\sqrt{\gamma}(R_0+\eta)$};
\end{tikzpicture}
}
\newcommand{\figC}{
\begin{tikzpicture}[scale=0.5, line cap=round,line join=round,>=triangle 45,x=1.0cm,y=1.0cm]
	\clip(0.6313244896864446,0.26698780636250397) rectangle (10.5108510279224,10.31427988958736);
	\draw [line width=2.pt,domain=0.6313244896864446:10.5108510279224] plot(\x,{(--12.-0.*\x)/4.});
	\draw [line width=2.pt] (3.,0.26698780636250397) -- (3.,10.31427988958736);
	\draw [line width=2.pt,domain=0.6313244896864446:10.5108510279224] plot(\x,{(--28.-0.*\x)/4.});
	\draw [line width=2.pt] (7.,0.26698780636250397) -- (7.,10.31427988958736);
	\draw [->,line width=1.2pt] (7.,5.) -- (8.,5.);
	\draw [->,line width=1.2pt] (5.,7.) -- (5.,8.);
	\draw (3.874791692805418,6.250625577633707) node[anchor=north west] {$\eta_{i,j}$};
	\draw (5.160994204387079,5.038985530491564) node[anchor=north west] {$b_{i,j}$};
	\draw (7.230102592583666,6.511594203172016) node[anchor=north west] {$\tilde{m}^1_{i+\frac{1}{2},j}$};
	\draw (4.862744346629013,9.41953031631316) node[anchor=north west] {$\tilde{m}^2_{i,j+\frac{1}{2}}$};
	\begin{scriptsize}
		\draw [color=black] (5.,5.)-- ++(-3.5pt,-3.5pt) -- ++(7.0pt,7.0pt) ++(-7.0pt,0) -- ++(7.0pt,-7.0pt);
	\end{scriptsize}
\end{tikzpicture}
}
\newcommand{\figB}{
\begin{tikzpicture}[line cap=round,line join=round,>=triangle 45,x=0.5cm,y=0.5cm]
	\clip(4.403327785518104,3.5831859932061794) rectangle (16.660271757042725,13.599613324774687);
	\draw (13.312676306755355,11.596327858460985) node[anchor=north west] {$Fv^1_{i+\frac{1}{2},j}$};
	\draw (5.826714827372576,7) node[anchor=north west] {$(x_F,y_F)$};
	\draw [line width=2.pt] (13.,13.)-- (13.,5.);
	\draw [line width=2.pt] (13.,5.)-- (5.,5.);
	\draw [line width=2.pt] (5.,5.)-- (5.,13.);
	\draw [line width=2.pt] (5.,13.)-- (13.,13.);
	\draw [line width=2.pt] (5.,9.)-- (13.,9.);
	\draw [line width=2.pt] (9.,13.)-- (9.,5.);
	\draw [line width=2.pt] (13.,11.)-- (10.580398465464366,9.);
	\draw [line width=2.pt] (10.580398465464366,9.)-- (9.,8.506700548266654);
	\draw [line width=2.pt] (9.,8.506700548266654)-- (6.708551585483659,8.136364927822294);
	\draw (6.222100116776596,7.853347118769595) node[anchor=north west] {$v^{1,fl}$};
	\begin{scriptsize}
		\draw [color=black] (6.708551585483659,8.136364927822294)-- ++(-4.5pt,-4.5pt) -- ++(9.0pt,9.0pt) ++(-9.0pt,0) -- ++(9.0pt,-9.0pt);
		\draw [color=black] (13.,11.) circle (4.5pt);
		\draw [color=ffqqqq] (10.580398465464366,9.) ++(-4.5pt,0 pt) -- ++(4.5pt,4.5pt)--++(4.5pt,-4.5pt)--++(-4.5pt,-4.5pt)--++(-4.5pt,4.5pt);
		\draw [color=ffqqqq] (9.,8.506700548266654) ++(-4.5pt,0 pt) -- ++(4.5pt,4.5pt)--++(4.5pt,-4.5pt)--++(-4.5pt,-4.5pt)--++(-4.5pt,4.5pt);
	\end{scriptsize}
\end{tikzpicture}
}

\newcommand\BibTeX{{\rmfamily B\kern-.05em \textsc{i\kern-.025em b}\kern-.08em
T\kern-.1667em\lower.7ex\hbox{E}\kern-.125emX}}

\usepackage{hyperref} 
\hypersetup{
    colorlinks=true,                          
    linkcolor=blue, % Couleur des liens internes
    citecolor=red, % Couleur des num�ros de la biblio dans le corps
    urlcolor=blue  } % Couleur des url
    
\newcommand{\x}{\mbf{x}}

\newcommand{\mbf}[1]{\mathbf{#1}}			%

\newcommand{\Ht}{\tilde{H}}
\newcommand{\mt}{\tilde{m}}
\newcommand{\half}{\frac{1}{2}}
\newcommand{\idx}[1]{{#1 }}
\newcommand{\CFL}{\textnormal{CFL}}

\newcommand{\bdm}{\begin{displaymath}}
\newcommand{\edm}{\end{displaymath}}

\newcommand{\bea}{\begin{eqnarray} }
\newcommand{\eea}{\end{eqnarray} }

\newfont{\numerikEleven}{ecrm1000}
\newfont{\numerikTen}{cmss10}
\newfont{\numerikNine}{cmss9}
\newfont{\numerikEight}{cmss8}

\makeatletter
\def\ps@pprintTitle{%
	\let\@oddhead\@empty
	\let\@evenhead\@empty
	\let\@oddfoot\@empty
	\let\@evenfoot\@empty}
\makeatother

\begin{document} 
%!=========================================================================
%!
%!      F R O N T    M A T T E R 
%!
\begin{frontmatter}
%-------------------------------------------------------
% TITLE
\title{A semi-implicit two dimensional solver for a covariant formulation of the shallow water equations } 
%-------------------------------------------------------
%-------------------------------------------------------
% AUTHORS
\author[UniVR]{Maurizio Tavelli}
\ead{maurizio.tavelli@univr.it}

\author[UniTN]{Olindo Zanotti}
\ead{olindo.zanotti@unitn.it}

\address[UniVR]{Department of Engineering for Innovation Medicine, University of Verona, Strada le Grazie 15, Verona, 37134, Italy}

\address[UniTN]{Laboratory of Applied Mathematics, DICAM, University of Trento, via Mesiano 77, 38123 Trento, Italy}

%-------------------------------------------------------
% INSTITUTIONS
%\address[UniTN]{Department of Civil, Environmental and Mechanical Engineering, 
%University of Trento, Via Mesiano 77, 38123 Trento, Italy.} 
%-------------------------------------------------------

%-------------------------------------------------------
% ABSTRACT
\begin{abstract} \color[rgb]{0,0,0}
In this paper we combine a flexible covariant formulation of the shallow water equations with the semi-implicit numerical scheme developed over the years by Casulli and collaborators. 
After adopting an orthogonal, but non-orthonormal, coordinate basis on two dimensional manifolds, and by writing the divergence of symmetric tensors in a way that avoids the introduction of Christoffel symbols, the shallow water equations preserve a very close resemblance to the usual one expressed in Cartesian coordinates.  
In this way, a stable semi-implicit scheme can be derived by using an implicit discretization for the gradient of surface
elevation in the momentum equations and for the velocity in the continuity equation, with stability properties that are independent of the celerity.
We have tested the new method over a variety of challenging benchmarks, including, among the others, the smooth wave propagation over a water globe and the deformation of an artery branch. 
Two appealing additional features make the method particularly powerful with respect to oceanographic applications: firstly, 
thanks to the wetting and drying ability of our semi-implicit approach,  
no pathological behaviors occur at the poles; secondly, the scheme is naturally well-balanced,  and  
it is able to preserve perfect stationarity, up to machined precision, 
of the entire ocean configuration of the earth.
\end{abstract}
%-------------------------------------------------------

%-------------------------------------------------------
% KEY WORDS
%-------------------------------------------------------
\end{frontmatter}
%!===================================================Partially======================

%-----------------------------------
% CONTENTS
%  This will deseapear in the submitted version
%\tableofcontents
%------------------------------------

%=========================================================================
%==========         I N T R O D U C T I O N
% 
\section{Introduction} 
\label{sec.introduction}
In spite of representing a crude simplification with respect to the Navier Stokes equations for fluid dynamics, the shallow water equations still attract a lot of interest, for their effectiveness in modeling various kinds of free surface  flows under quite different  physical and morphological conditions. 
The scientific literature on the subject is actually so large that it is virtually impossible to cover in a few sentences the
advancements that have been performed, both with respect to the variety of applications and with respect to the numerical schemes that have been developed.
We just mention that a very prominent application is represented by tzunami wave propagation, for which  
high order numerical schemes were proposed by \cite{CASTRO2012}. 
We address the interested reader to a few excellent reviews of this topic, such as those presented by
\cite{Casulli2022,Delis2021,GarciaNavarro2019,Toro2024,Ngatcha2024}.

Among the large family of numerical schemes that have been developed over the years for the solution of the shallow water equations, semi-implicit schemes on staggered grids represent a notable case. They were first developed on Cartesian meshes by \cite{Casulli1990} and then extended to unstructured ones in \cite{Casulli2000}. The shallow water system is characterized by the coexistence of several different scales. 
Pressure waves travel at a speed given by the celerity $\sqrt{gH}$, which can become considerably large depending on the value of the total water height $H$.
The advective part, on the contrary, is governed by the fluid velocity, which is typically much smaller. 
In these conditions, a standard explicit scheme would be limited by a sever $\CFL$ condition \citep{Courant1928} driven 
by the fastest wave speeds, i.e. the celerity. On the contrary,
semi-implicit methods on staggered meshes perform an implicit discretization of a few crucial 
quantities that are carefully selected to obtain a simple non linear system to be solved at each time iteration. More specifically, \cite{Casulli1990} showed that it is possible to obtain
a simple non-linear system whose non-linearities appear only on the diagonal, while 
the linear part is provably symmetric
and positive semi-definite. 
Over the years, this methodology has been extended  both to more advanced high order methods \citep{Dumbser2013, Tavelli2014} and to challenging PDE systems
\citep{Brugnano2009,Casulli2012,Tavelli2013,Lucca2023, Lucca2025, Dumbser2026}. 

When the shallow water equations need to be solved on curved geometries, like in oceanographic applications where the earth curvature cannot be neglected \citep{Kolar1994,SATOSHI2010,ZEITLIN2007,RINGLER2010}, or even in blood flows within arteries, curvilinear coordinates must be adopted. 
In this respect, the mathematical tools borrowed from differential geometry turn out to be very convenient, and  covariant formulations of the shallow water equations become necessary. Several promising attempts have 
already been proposed, such as, 
among the others, those of \cite{Kolar1994}, \cite{WINTERMEYER2017}, \cite{Arpaia2020}, \cite{Carlino2023}, \cite{Montoya2026}, where
a variety of discretization techniques have been explored.

Our approach for the solution of the shallow water equations on curved manifolds is based on two key features.
Firstly, we adopt a particularly simple covariant version of the shallow water equations in two space dimensions,  which fully exploits the tensor expression of the divergence operator. The net result of this simple gimmick is that no Christoffel symbols appear in the equations, which therefore maintain the closest possible resemblance of their Cartesian counterpart. As a side effect, the evolved quantities acquire the geometric factor $\sqrt{\gamma}$ ($\gamma$ being the determinant of the curved manifold in the specific coordinates adopted), while an extra algebraic source term appears on the right hand side.
When the coordinates are Cartesian, $\gamma=1$, covariant and contra-variant components of any tensor coincide,   
and the covariant form of the equations naturally reduces to the usual expressions. Secondly, we extend the semi-implicit schemes introduced  by \cite{Casulli1990,Casulli2000} to this new formulation, preserving all of their numerical advantageous properties. In particular, 
the semi-implicit approach, 
removes the pathological effects due to coordinate singularities, such as those affecting the poles on a spherical surface.
Moreover, the numerical scheme is also naturally well balanced, in the sense of ~\cite{Castro2008,Castro2017,Castro2017Book,castro2020well}, as already proved by  \cite{BOSCHERI2023}.

We have validated the new approach over a number of crucial tests for the classical shallow water equations, including: the propagation of a smooth wave over waterland across the poles, the well balanced property of the entire earth oceans, 
two standard Riemann problem,
the simulation of a steady state geostrophic flow and the simulation of blood flow in an artery with deformation.

The plan of the paper is the following: in Sect.~\ref{sec.GHeqs} we present the governing equations of shallow water written in a particular simple covariant form that avoids Christoffel symbols.
Sect.~\ref{sec.schemes} is instead devoted to the presentation of the semi-implicit numerical scheme. Sect.~\ref{sec.tests} contains the  numerical results of our investigation, and Sect.~\ref{sec.conclusions} concludes our work. Regarding the notation,
we make an important distinction among tensor indices and discretization indices:
\begin{itemize}
	\item Indices {\emph{i}} and  {\emph{j}} are {\emph{only}} used as spatial discretization indices. appearing as subscripts.
	\item Index {\emph{n}} is {\emph{only}} used as temporal discretization index. appearing as superscript.
	\item Greek indices {\emph{$\alpha$}}, {\emph{$\beta$}}, {\emph{$\mu$}} and  {\emph{$\nu$}} are {\emph{only}} used as tensor indices, ranging from 1 to 2, appearing either covariant or contra-variant.
\end{itemize}
Moreover,
we adopt the standard Einstein summation convention over  repeated (tensor) indices. 
Finally we adopt the standard IS system of units all along the paper. In this context, especially on very large scales, it is critical to adopt a proper normalization of the involved equations, as detailed in Section \ref{sec.Scaling}.

%======================================================================
\section{The covariant formulation of the Shallow Water equations}
\label{sec.GHeqs}
\subsection{Choice of the curvilinear coordinates}
We plan to solve the shallow water equations written in covariant form, assuming that the acceleration vector $\vec g$ is locally perpendicular to a two-dimensional manifold. The manifold itself is covered by a single coordinate chart, and the coordinates are denoted as $x^\alpha$, with $\alpha=1,2$. Using more than one chart is in principle possible but
we will not consider this case here.
The covariant spatial metric is given by $\gamma_{\alpha\beta}$, which can be represented as a $2\times 2$ symmetric matrix, which allows to compute physical distances on the manifold from
\begin{equation}
	\label{eq:metric}
	d\ell^2= \gamma_{\alpha\beta}dx^\alpha dx^\beta=\gamma_{11}(dx^1)^2+\gamma_{22}(dx^2)^2 \,, 
\end{equation}
where we have assumed, as it will be done hereafter, that the metric is diagonal.
Since a natural application of the shallow water equations occur on  the surface of the earth, two-dimensional spherical coordinates will be often adopted, hence with $x^1=\theta$, $x^2=\phi$. In that case we have\footnote{Note that $\theta$ is measured from the vertical axis, hence it is a polar angle.}
\begin{eqnarray}
	\label{eq:SphericalCoordinates-1}
	X(\theta,\phi)&=&R\sin\theta \cos\phi\\
	\label{eq:SphericalCoordinates-2}
	Y(\theta,\phi)&=&R\sin\theta \sin\phi\\
	\label{eq:SphericalCoordinates-3}
	Z(\theta,\phi)&=&R\cos\theta \,,
\end{eqnarray}
where $R$ is the radius of the sphere, and the spatial metric becomes simply
\begin{equation}
	\gamma_{\alpha\beta }(\theta,\phi,R)=
	\begin{pmatrix}
		R^{2} & 0\\[6pt]
		0        & R^{2}\sin^{2}\theta
	\end{pmatrix}\,,\hspace{1cm}\gamma=det(\gamma_{\alpha\beta})\,,
	\label{eq:metric_spherical}
\end{equation}
with $\sqrt{\gamma}=R^2\sin \theta$. An alternative example of the metric in cylindrical coordinates, suitable for applications to arterial flows, will be shown in Sect.~\ref{sec.Tube}.
%=============================================================================
\subsection{Formulation of the equations}
The shallow water equations  on a curved manifold can be obtained starting from the usual continuity and momentum equations 
\begin{eqnarray}
	\label{eq:cov-1}
	&&\frac{\partial \rho}{\partial t}+\nabla_\alpha (\rho u^\alpha)=0\,\\
	\label{eq:cov-2}
	&&\frac{\partial u^\alpha}{\partial t}+ u^\beta  \nabla_\beta u^\alpha+ \nabla_\beta T^{\alpha\beta}=0\,,\hspace{1cm}(\alpha,\beta=1,2)
\end{eqnarray}
where  
$T^{\alpha\beta}$ are the components of the standard stress tensor for a perfect fluid, given by [see Eq. (26.8) by \cite{Mihalas1984}], i.e.
\begin{equation}
	T^{\alpha\beta}=p \gamma^{\alpha\beta}\,.
\end{equation}
Now, we first introduce the total water depth 
$H=\eta-b$, where $\eta$ is the water surface elevation and $b$ the prescribed bathymetry.
Then, recalling the usual assumptions of the shallow water approximation, namely:
\begin{itemize}
	\item Incompressibility: $\nabla_\beta u^\beta=0$\,;
	\item Integration along the local normal to the manifold (henceforth referred to as \emph{vertical averaging}) of any quantity along the $z$ direction, such that, for instance, $v^\alpha=\frac{1}{H}\int_{b}^\eta u^\alpha\,dz$  \,;
	\item Hydrostatic vertical equilibrium, such that $p=p_{e}+ g (\eta-z)$, where $p$ has been normalized by the constant $\rho$, 
	and where $p_{e}$ is the external pressure,
\end{itemize}
it is possible to show that Eq.~\eqref{eq:cov-1}-\eqref{eq:cov-2} transform into (see \ref{sec:appendixA} for the derivation)
\begin{eqnarray}
		\label{eq:CSW-1}
	&&\frac{\partial (\sqrt{\gamma}H)}{\partial t}+\partial_\beta \left(\sqrt{\gamma}m^\beta\right)=0\,,\\
		\label{eq:CSW-2}
	&&\frac{\partial (\sqrt{\gamma}m_\alpha)}{\partial t}+\partial_\beta \left(\sqrt{\gamma}\frac{m_\alpha m^\beta}{H}\right)+g\sqrt{\gamma}H\partial_\alpha \eta=\frac{1}{2}\sqrt\gamma\left[\frac{m^\mu m^\nu}{H}\partial_\alpha\gamma_{\mu\nu}
%	-H^2\partial_\alpha g
	\right]\,,
\end{eqnarray}
where we have defined  the mass flux as $m^\alpha=H v^\alpha$.
%One can arrive to the same final form following the approach of  \cite{Baldauf2020,Carlino2023}. 
%Note that, for a general treatment 
%aimed at a proper modeling of tides, we have allowed the
%gravity field to have a spatial, and also temporal, dependence, i.e. $g=g(x^\alpha,t)$.
Note that one has to distinguish among the coordinate velocity $u^\alpha=dx^\alpha/dt$ and the physical velocity $|\vec u|=\sqrt{\gamma_{\alpha\beta}u^\alpha u^\beta}$.
%Using Eq.~\eqref{eq:metric} one can also write the single component as
%$d\ell_\alpha /dt=\sqrt{\gamma_{\alpha\beta}}dx^\beta/dt=\sqrt{\gamma_{\alpha\beta}}v^\beta$.
For convenience we may write $\Ht=\sqrt{\gamma}H$ and $\mt_\alpha=\sqrt{\gamma}m_\alpha$, and we emphasize that, even if $\eta$ may assume values smaller than $b$, the total water depth is in general a non linear function of $\eta$, namely: $H(\eta)=\max(0,\eta-b)$. Hence, a non linearity is  introduced in order to allow for wetting (when $\eta>b$) and also for drying (when $\eta<b$). In this last case we simply obtain $H(\eta)=0$ and no momentum equation need to be solved, i.e. $m_\alpha=0$. See also \cite{Casulli2009}.
The previous system then reads
\begin{eqnarray}
	\label{eq:CSW-1-f}
	&&\partial_t \Ht+\partial_\beta \mt^\beta=0\,,\\
	\label{eq:CSW-2-f}
	&&\partial_t \mt_\alpha+\partial_\beta \left(\mt_\alpha v^\beta \right)+\underbrace{g\Ht\partial_\alpha \eta}_{\text{\normalfont pressure gradients}}=\underbrace{\frac{1}{2}\mt^\mu v^\nu\partial_\alpha \gamma_{\mu\nu}}_{\text{\normalfont curvature correction}}\,.
%	-\underbrace{\frac{1}{2\sqrt{\gamma}}\Ht^2\partial_\alpha g}_{\text{\normalfont gravity gradients}}\,.
\end{eqnarray}
%
%\begin{eqnarray}
%\partial_j \left(\sqrt{\gamma} \frac{1}{2}g H^2 \delta_i^{j}\right)=\partial_i \left(\sqrt{\gamma} \frac{1}{2}g H^2 \right)= \frac{1}{2} g \partial_i \left(\sqrt{\gamma}H^2 \right)+  \frac{1}{2} \sqrt{\gamma}H^2 \partial_i g
%\end{eqnarray}
If we compare Eq.~\eqref{eq:CSW-1-f}-\eqref{eq:CSW-2-f} above with Eq.~(11) by \cite{Casulli2022} we can see that the two expressions are formally very close to each other. As a matter of fact, the curvature effects are entirely accounted for 
by the first term on the right hand side of \eqref{eq:CSW-2-f}, by the factor $\sqrt{\gamma}$, 
as well as by the crucial distinction among covariant and contravariant vector components. 
For example, while in the continuity equation~\eqref{eq:CSW-1-f}
the momentum field $\tilde m^\beta$ enters as a contravariant vector, 
in the momentum equation \eqref{eq:CSW-1-f} it appears as a covariant one.
We change from one form to the other through the metric as
\begin{eqnarray}
	\mt^\mu=\gamma^{\mu\nu}\mt_\nu.
\end{eqnarray}
Since the metric is diagonal, this amounts to
\begin{eqnarray}
	\mt^{1}=\gamma^{11}\mt_1=\frac{\mt_1}{\gamma_{11}}\,, \qquad \mt^{2}=\gamma^{22}\mt_2=\frac{\mt_2}{\gamma_{22}} \,.
	\label{eq:cotoconta}
\end{eqnarray}
%
%We also stress that,  when $g$ is not constant, the equilibrium state (i.e. $v^\alpha\equiv 0$) is not given by $\partial_\alpha \eta=0$ but rather by $g \partial_\alpha \eta=-\frac{1}{2}H\partial_\alpha g$ that is responsible for  the tidal phenomenon. 

%=============================================================================
\subsection{Inclusion of the Coriolis force}
\label{sec:Coriolis}
For rotating systems like the earth, it is necessary to include also the non-inertial effects due to the Coriolis force. In vector form, the Coriolis acceleration in the corotating frame is 
$\vec a_C = -2\vec\Omega \times \vec v$. In components, this can be written as:
\begin{equation}
	(a_C)^\alpha=-2\epsilon^{\alpha\beta\gamma}\Omega_\beta v_\gamma \,,
\end{equation} 
where $\epsilon^{\alpha\beta\gamma}$ is the Levi-Civita tensor, which has components $\epsilon^{\alpha\beta\gamma}=1/\sqrt{\gamma}\,\cdot [\alpha\beta\gamma]$, and where $[\alpha\beta\gamma]$ are the totally antisymmetric symbols. Having in mind applications on the earth surface using coordinates $(\theta,\phi)$, the two relevant components of the Coriolis acceleration are
\begin{align}
	(a_C)^\theta&=\frac{2}{\sqrt{\gamma}}\Omega_r v_\phi=\frac{2}{\sqrt{\gamma}}\cos\theta\,\Omega\, v_\phi\\
	(a_C)^\phi&=-\frac{2}{\sqrt{\gamma}}\Omega_r v_\theta=-\frac{2}{\sqrt{\gamma}}\cos\theta\,\Omega\, v_\theta\,,
\end{align}	
where the angular velocity vector $\vec\Omega$ is oriented along the $z$ axis. As shown by \cite{Casulli2000}, Coriolis terms ought to be treated implicitly. This means that, if we focus just on Coriolis corrections to the momentum equation \eqref{eq:CSW-2-f}, and after introducing the \emph{Coriolis parameter} as $f=2\cos\theta\Omega$, the numerical discretization of the Coriolis correction will be
\begin{align}
	\label{Coriolis-1}
	& v^{1,n+1}=v^{1,fl}+ \Delta t \frac{f}{\sqrt{\gamma}} v_2^{n+1}=v^{1,fl}+ \Delta t \frac{f}{\sqrt{\gamma_{11}}\sqrt{\gamma_{22}}} \gamma_{22} v^{2,n+1} \\
	\label{Coriolis-2}
	& v^{2,n+1}=v^{2,fl}- \Delta t \frac{f}{\sqrt{\gamma}} v_1^{n+1}
	=v^{2,fl}- \Delta t \frac{f}{\sqrt{\gamma_{11}}\sqrt{\gamma_{22}}} \gamma_{11} v^{1,n+1}
\end{align}
which amounts to
\begin{align}
	\label{Coriolis-1a}
	& v^{1,n+1}=v^{1,fl}+ \Delta t f \sqrt{\frac{\gamma_{22}}{\gamma_{11}}} v^{2,n+1} \\
	\label{Coriolis-2b}
	& v^{2,n+1}=v^{2,fl}- \Delta t f \sqrt{\frac{\gamma_{11}}{\gamma_{22}}} v^{1,n+1}\,,
\end{align}
where $v^{1,fl}$ and $v^{2,fl}$ are the components of the advection contribution, see Section \ref{sec.Fm}. Note that 
\eqref{Coriolis-1a}-\eqref{Coriolis-2b} collapse
to the same expression found by \cite{Casulli2000} if an Euclidean metric is considered. 
The previous system can easily be inverted in the unknowns $v^{\alpha,n+1}$ and reads
\begin{eqnarray}
	v^{1,n+1}&=&\frac{1}{1+\Delta t^2 f^2}\left(v^{1,fl}+f\Delta t \sqrt{\frac{\gamma_{22}}{\gamma_{11}}}v^{2,fl}  \right) \\
	v^{2,n+1}&=&\frac{1}{1+\Delta t^2 f^2}\left(v^{2,fl}-f\Delta t \sqrt{\frac{\gamma_{11}}{\gamma_{22}}}v^{1,fl}  \right).
\end{eqnarray}
%======================================================================
\section{Semi implicit numerical scheme}
\label{sec.schemes}
\subsection{Staggered meshes for the reference space}
We discretize a reference two dimensional domain $\Omega_R$ using a simple uniform mesh. As stated above, we then assume that the metric tensor $\gamma_{\alpha\beta}$ is diagonal, meaning that the coordinate basis is orthogonal. 
%In general we may relax this condition, however the resulting mesh should be orthogonal with respect to the coordinates. 
This is an extension of the concept of unstructured orthogonal meshes introduced by \cite{Casulli2000}. See also \cite{Tavelli2014}.
%In this context the coordinate system integral to the edges should be orthogonal. 
For convenience of notation, and in order to help the comparison with already existing algorithms, in this Section we
assume $x^1=x$ and $x^2=y$. 
We cover the coordinate space $\Omega_R$ with $N_x \times N_y$ segments in the $x$ and $y$ direction, respectively. Each center is identified with the coordinates $x_\idx{i}$ and $y_\idx{j}$ for $i=1\ldots N_x$,  $j=1\ldots N_y$. We then use two edge-based \emph{staggered grids} that are identified with the half indices, namely the center of any vertical edge is written as $(x_{\idx{i+\half}},y_\idx{j})$ and $(x_\idx{i}, y_\idx{j+\half})$ for the horizontal faces.  We then call $\Delta x = x_\idx{i+\half}-x_\idx{i-\half}$ and $\Delta y = y_\idx{j+\half}-y_\idx{j-\half}$, which are both constant.
\subsection{Numerical approximation}
As originally proposed by \cite{Casulli1990}, the vertically averaged velocity is discretized over the edges while the free surface elevation, the bathymetry  and the total water depth are formally defined over the centers. 
This is often referred to as the \emph{primary} discretization. 
According to the notation specified at the end of Sect.~\ref{sec.introduction}, we use $i$ and $j$ as spatial discretization indices,
$n$ as a temporal discretization index, and Greek letters as tensorial indices.
After defining the solution time $t^\idx{n}=t^{\idx{n-1}}+\Delta t^{\idx{n}}$ as a function of the time interval $\Delta t^n$, the corresponding primary numerical quantities are  indicated as follows:
\begin{align}
	&\eta_\idx{i,j}^\idx{n}=\eta\left(x_\idx{i},y_\idx{j}, t^\idx{n}\right), \qquad b_\idx{i,j}=b\left(x_\idx{i},y_\idx{j}\right), \qquad 	\Ht_\idx{i,j}\left(\eta_\idx{i,j}^\idx{n}\right)=\sqrt{\gamma}_\idx{i,j} max\left(0, \eta_\idx{i,j}^\idx{n}-b_\idx{i,j}\right)\,, \\
 &\mt^{1,\idx{n}}_\idx{i+\half,j}=\mt^1\left(x_\idx{i+\half},y_\idx{j}, t^\idx{n}\right), 
 \qquad
 \mt^{2,\idx{n}}_\idx{i,j+\half}=\mt^2\left(x_\idx{i},y_\idx{j+\half}, t^\idx{n}\right), 
 \qquad
	\gamma^{\mu\nu}_\idx{i,j}=\gamma^{\mu\nu}\left(x_\idx{i},y_\idx{j}\right)\,.
	\label{eq:variables_1}
	% \nonumber% \\ %(\sqrt{\gamma})^{rl}_\idx{i,j}=(\sqrt{\gamma})^{rl}\left(x_\idx{i},y_\idx{j}\right)
\end{align}
All these quantities are assumed to be constant over the elements and edges where they are defined, while the velocity is perpendicular to the edge where it is defined.
\begin{figure}[!h]
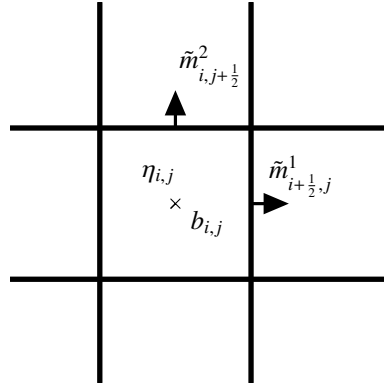

	\centering
	\figC
	\caption{Schematic view of the adopted discretization.}
	\label{fig:var}
\end{figure}
A schematic view of the adopted discretization is reported in Fig \ref{fig:var}. 
Analogously, we use the term \emph{dual} discretization to indicate a discretization that is staggered with respect to the primary one. Hence, we compute the corresponding dual components as:
\begin{align}
&\eta_\idx{i+\half,j}^\idx{n}=  max\left( \eta_\idx{i,j}^\idx{n},\eta_\idx{i+1,j}^\idx{n} \right), \qquad
\eta_\idx{i,j+\half}^\idx{n}=  max\left( \eta_\idx{i,j}^\idx{n},\eta_\idx{i,j+1}^\idx{n} \right),  \\
&	b_\idx{i+\half,j}=  max\left( b_\idx{i,j},b_\idx{i+1,j} \right), \qquad
	b_\idx{i,j+\half}=  max\left( b_\idx{i,j},b_\idx{i,j+1} \right),  \\
&	\gamma^{\mu\nu}_\idx{i+\half,j}=  \gamma^{\mu\nu}\left(x_\idx{i+\half},y_\idx{j}\right), \qquad
	\gamma^{\mu\nu}_\idx{i,j+\half}=  \gamma^{\mu\nu}\left(x_\idx{i},y_\idx{j+\half}\right), \\	
&	\Ht\left(\eta_\idx{i+\half,j}^\idx{n}\right)=\sqrt{\gamma}_\idx{i+\half,j} max\left(0, \eta_\idx{i+\half,j}^\idx{n}-b_\idx{i+\half,j}\right), \qquad
	\Ht\left(\eta_\idx{i,j+\half}^\idx{n}\right)=\sqrt{\gamma}_\idx{i,j+\half} max\left(0, \eta_\idx{i,j+\half}^\idx{n}-b_\idx{i,j+\half}\right).
\label{eq:variable_dual_1}
\end{align}
A consistent semi-implicit approximation of the continuity equation \eqref{eq:CSW-1-f} reads
\begin{eqnarray}
	\Ht_\idx{i,j}^\idx{n+1}=\Ht_\idx{i,j}^\idx{n}
	-\Delta t\frac{\mt^{1,\idx{n+1}}_\idx{i+\half,j}-\mt^{1,\idx{n+1}}_\idx{i-\half,j}}{\Delta x}
	-\Delta t\frac{\mt^{2,\idx{n+1}}_\idx{i,j+\half}-\mt^{2,\idx{n+1}}_\idx{i,j-\half}}{\Delta y}\,,
	\label{eq:cont_disc_1}
\end{eqnarray}
while a finite difference approximation of the momentum equation reads
\begin{eqnarray}
	\label{eq:mom_disc_1-a}
	\mt_{1,\idx{i+\half,j}}^\idx{n+1}&=&F_{1,\idx{i+\half,j}}^\idx{n}-g_\idx{i+\half,j}^\idx{n+1}\Delta t \Ht_\idx{i+\half,j}^\idx{n}\frac{\eta_\idx{i+1,j}^\idx{n+1}-\eta_\idx{i,j}^\idx{n+1}}{\Delta x}  \\
	\mt_{2,\idx{i,j+\half}}^\idx{n+1}&=&F_{2,\idx{i,j+\half}}^\idx{n}-g_\idx{i,j+\half}^\idx{n+1} \Delta t \Ht_\idx{i,j+\half}^\idx{n}\frac{\eta_\idx{i,j+1}^\idx{n+1}-\eta_\idx{i,j}^\idx{n+1}}{\Delta y}.
	\label{eq:mom_disc_1-b}
\end{eqnarray}
Here $F_1^\idx{n}$ and $F_2^\idx{n}$ are stable approximations of the non-linear convective, Coriolis, and distortion terms. Further forces such as gravity gradients may be added. Since in a spherical coordinate system the distortion close to the poles is very large (due to the factor $\sqrt{\gamma}$), then the velocity field in the reference space can be huge. Hence an explicit discretization of those terms may be affected from a severe $\CFL$ time step restriction for some configurations. In Sect.~\ref{sec.Fm}, a semi-Lagrangian approach, following an idea originally proposed by \cite{Stelling2003}, is presented to cope with this potential difficulty. 
In Eq.~\eqref{eq:mom_disc_1-a}-\eqref{eq:mom_disc_1-b} the free surface is discretized implicitly in order to avoid a time restriction driven by the celerity, while the terms $\Ht$ are discretized explicitly to avoid off-diagonal non-linear terms.
This approach has been widely adopted after \cite{Casulli1990}. 
After applying the transformation Eq. \eqref{eq:cotoconta} to the discrete momentum \eqref{eq:mom_disc_1-a}-\eqref{eq:mom_disc_1-b}, we replace the obtained quantities in the discrete continuity equation, obtaining a system with the free surface elevation as the only unknown, i.e.
\begin{eqnarray}
	\Ht_\idx{i,j}^\idx{n+1}
	%&=&\Ht_\idx{i,j}^\idx{n}
	%-\frac{\Delta t}{\Delta x}\left[\mt^{1,\idx{n+1}}_\idx{i+\half,j}-\mt^{1,\idx{n+1}}_\idx{i-\half,j}\right]
	%-\frac{\Delta t}{\Delta y}\left[\mt^{2,\idx{n+1}}_\idx{i,j+\half}-\mt^{2,\idx{n+1}}_\idx{i,j-\half}\right] \nonumber \\
	%&=& -\frac{\Delta t}{\Delta x}\left[\gamma^{11}_{\idx{i+\half,j}}Fm^{\idx{n+1}}_{1,\idx{i+\half,j}}
	%                            -\gamma^{11}_{\idx{i-\half,j}}Fm^{\idx{n+1}}_{1,\idx{i-\half,j}}\right]
	%    -\frac{\Delta t}{\Delta y}\left[\gamma^{22}_{\idx{i,j+\half}}Fm^{\idx{n+1}}_{2,\idx{i,j+\half}}
	%                            -\gamma^{22}_{\idx{i,j-\half}}Fm^{\idx{n+1}}_{2,\idx{i,j-\half}}\right] \nonumber \\
	&=& \Ht_\idx{i,j}^\idx{n}-\frac{\Delta t}{\Delta x}\left[F^{1,\idx{n}}_{\idx{i+\half,j}}-F^{1,\idx{n}}_{\idx{i-\half,j}}\right]
        -\frac{\Delta t}{\Delta y}\left[F^{2,\idx{n}}_{\idx{i,j+\half}}-F^{2,\idx{n}}_{\idx{i,j-\half}}\right] \nonumber \\
     &&+\frac{\Delta t^2}{\Delta x^2}\left[ 
     	\gamma^{11}_{\idx{i+\half,j}}g_\idx{i+\half,j}^\idx{n+1} \Ht_\idx{i+\half,j}^\idx{n}\left(\eta_\idx{i+1,j}^\idx{n+1}-\eta_\idx{i,j}^\idx{n+1}\right)
       -\gamma^{11}_{\idx{i-\half,j}}g_\idx{i-\half,j}^\idx{n+1} \Ht_\idx{i-\half,j}^\idx{n}\left(\eta_\idx{i,j}^\idx{n+1}-\eta_\idx{i-1,j}^\idx{n+1}\right)
       \right] \nonumber \\
     &&+\frac{\Delta t^2}{\Delta y^2}\left[ 
     \gamma^{22}_{\idx{i,j+\half}}g_\idx{i,j+\half}^\idx{n+1} \Ht_\idx{i,j+\half}^\idx{n}\left(\eta_\idx{i,j+1}^\idx{n+1}-\eta_\idx{i,j}^\idx{n+1}\right)
    -\gamma^{22}_{\idx{i,j-\half}}g_\idx{i,j-\half}^\idx{n+1} \Ht_\idx{i,j-\half}^\idx{n}\left(\eta_\idx{i,j}^\idx{n+1}-\eta_\idx{i,j-1}^\idx{n+1}\right)
     \right]. 
	\label{eq:eta_eqn_1}
\end{eqnarray}
The previous system can be written in a compact form as
\begin{eqnarray}
	\Ht_\idx{i,j}\left(\eta^\idx{n+1}_\idx{i,j},b_\idx{i,j}\right)+S^\idx{n}_\idx{i,j}\eta^\idx{n+1}_\idx{i,j}-s^\idx{n}_\idx{i+\half,j}\eta^\idx{n+1}_\idx{i+1,j}-s^\idx{n}_\idx{i-\half,j}\eta^\idx{n+1}_\idx{i-1,j}-s^\idx{n}_\idx{i,j+\half}\eta^\idx{n+1}_\idx{i,j+1}-s^\idx{n}_\idx{i,j-\half}\eta_\idx{i,j-1}^\idx{n+1}=b^\idx{n}_\idx{i,j},
	\label{eq:eta_eqn_2}
\end{eqnarray}
where
\begin{eqnarray}
	s^\idx{n}_\idx{i\pm\half,j}&=&\frac{\Delta t^2}{\Delta x^2}\gamma^{11}_{\idx{i\pm\half,j}}g_\idx{i\pm\half,j}^\idx{n+1}\Ht_\idx{i\pm\half,j}^\idx{n}\,, \\
	s^\idx{n}_\idx{i,j\pm\half}&=&\frac{\Delta t^2}{\Delta y^2}\gamma^{22}_{\idx{i,j\pm\half}}g_\idx{i,j\pm\half}^\idx{n+1}\Ht_\idx{i,j\pm\half}^\idx{n}\,, \\
	S^\idx{n}_\idx{i,j} &=& s^\idx{n}_\idx{i+\half,j}+s^\idx{n}_\idx{i-\half,j}+ s^\idx{n}_\idx{i,j+\half}+s^\idx{n}_\idx{i,j-\half}\,, \\
	b^\idx{n}_\idx{i,j} &=& \Ht_\idx{i,j}^\idx{n}-\frac{\Delta t}{\Delta x}\left[F^{1,\idx{n}}_{\idx{i+\half,j}}-F^{1,\idx{n}}_{\idx{i-\half,j}}\right]
	-\frac{\Delta t}{\Delta y}\left[F^{2,\idx{n}}_{\idx{i,j+\half}}-F^{2,\idx{n}}_{\idx{i,j-\half}}\right]\,. 
\end{eqnarray}
It is clear from \eqref{eq:eta_eqn_2} that the linear part is symmetric and diagonally dominant, hence semi-positive definite. Furthermore, the resulting stencil  is composed by at most five non zero elements, which is very convenient from a computational point of view. This is valid irrespective of $\gamma^{\alpha\beta}$, but simply due to its structure. 
Hence, the previous system can be rewritten in the following even more compact form
\begin{eqnarray}
	\Ht(\eta)+T\eta=b,
\end{eqnarray}
that is nothing but a weakly non-linear system. This kind of system can be solved using an efficient Newton algorithm, whose convergence is guaranteed in a finite number of iterations,  \citep{Brugnano2009}.  In summary, we are interested in the value of $\eta$ such that $f(\eta)=\Ht(\eta)+T\eta-b=0$. 
Starting from $\xi^{(0)}>b$ we set the iteration stage as
\begin{eqnarray}
	\xi^{(l+1)}=\xi^{(l)}-\left[G^{(l)}+T\right]^{-1}\left[\Ht(\xi^{(l)})+T\xi^{(l)}-b\right] \qquad l=0,1,2,\ldots
\end{eqnarray} 
where 
\begin{eqnarray}
	G^{(l)}_\idx{i,j}=\left.\frac{d\Ht}{d\xi}\right|_{i,j}=\left\{
	\begin{array}{lc}
		\sqrt{\gamma}_\idx{i,j} & \mbox{if } \xi_\idx{i,j}^{(l)}>b_\idx{i,j} \\
		0						& \mbox{otherwise,}
	\end{array}\right.
\end{eqnarray}
The stopping criteria is then $|f(\xi^{(l)})|<\epsilon$ where $\epsilon$ can be chosen as small as $\epsilon=10^{-10}$.
This algorithm implicitly includes and solves the non linearities associated to the wetting and drying, that is automatically embedded in the implicit solver. An alternative high order approach to solve this problem for the shallow water system was proposed by \cite{XING2010}.

\paragraph{Comments and remarks:} The quantities $\sqrt{\gamma}$ and $\gamma_{\alpha\beta}$ may become function of $\eta$ when $\eta$ is dimensionally comparable to the characteristic geometrical size of the manifold.
In this case we can  consider $\gamma_{\alpha\beta}$ and $\sqrt{\gamma}$ as the vertically averaged quantities using the trapezoidal rule, i.e. 
\begin{equation}
\sqrt{\gamma}\approx\frac{1}{2}\left[(\sqrt{\gamma})(\eta^\idx{n})+(\sqrt{\gamma})(b)\right]. 
\end{equation}
Concerning the term $(\sqrt{\gamma})(\xi)$ appearing in $H(\xi)$ and its Jacobian $G^{(l)}$ it may be written as 
\begin{eqnarray}
	G^{(l)}_\idx{i,j}&=&\left.\frac{d\Ht}{d\xi}\right|_{i,j}=\left.\frac{d}{d\xi}(\sqrt{\gamma}(\xi)\max(0,\xi-b))\right|_{i,j} %\nonumber \\
	=\left\{
 \begin{array}{lc}
 	\sqrt{\gamma}_\idx{i,j}+\left(\xi_{\idx{i,j}}-b_{\idx{i,j}}\right)\frac{d\sqrt{\gamma}}{d\xi} & \mbox{if } \xi_\idx{i,j}^{(l)}>b_\idx{i,j} \\
 	0						& \mbox{otherwise,}
 \end{array}\right.
 %\frac{1}{2}\left[(\sqrt{\gamma})\left(b_{\idx{i,j}}\right)+(\sqrt{\gamma})\left(\xi_{\idx{i,j}}\right) \right]+\left(\xi_{\idx{i,j}}-b_{\idx{i,j}}\right)\frac{d(\sqrt{\gamma})}{d\xi}=
 \label{eq:Ggammaeta}
\end{eqnarray}
that is a simple linearization of $\sqrt{\gamma}$. Since $\sqrt{\gamma}$ is a known function we can also use its analytical expression to better  approximate $G^{(l)}_{\idx{i,j}}$.
Note also that, even if $\Delta x$ and $\Delta y$ are constant in the reference space, their contribution in the non-linear system is not constant when $\gamma$ becomes a function of the coordinates. In the spherical coordinate system, for example, this would lead to larger velocities in the coordinate space with respect to the physical one.
%===================================================
\subsection{Pixel resolution}
We want now to extend the algorithm using sugbrids for the bathymetry and the fluxes, following \cite{Casulli2019}. The idea is that, even if the free surface elevation can be approximated using large cells, the bathymetry needs to be approximated with a finer mesh. In a Cartesian framework we then define a new quantity
\begin{eqnarray}
	\tilde{b}_\idx{i,j,i_s,j_s}:= b\left(x_{i-\half}+(i_s-0.5)\frac{\Delta x}{N^s_x},y_{j-\half}+(j_s-0.5)\frac{\Delta y}{N^s_y}\right) \qquad \forall i_s=1\ldots N_x^s \,\, j_s=1 \ldots N_y^s\,, %\,\, i=1\ldots N_x \,\, j=1\ldots N_y,
\end{eqnarray}
where $N_x^s$ and $N_y^s$ are the number of sub-intervals inside each computational cell. For simplicity we assume these quantities constant for every element. Of course if $N_x^s=N_y^s=1$ then the definition of $\tilde{b}$ coincides with the one adopted in \eqref{eq:variables_1}. We then need to generalize all the fluxes in \eqref{eq:variable_dual_1} and the volumes according to this new definition 

\begin{eqnarray}
	\tilde{b}_\idx{i+\half,j,j_s}&=&  max\left( \tilde{b}_\idx{i,j,N_x^s,j_s},\tilde{b}_\idx{i+1,j,1,j_s} \right),  \\
	\tilde{b}_\idx{i,j+\half,i_s}&=&  max\left( b_\idx{i,j,i_s,N_y^s},b_\idx{i,j+1,i_s,1} \right),  \\	
	\Ht_\idx{i,j}\left(\eta_\idx{i,j}^\idx{n}\right)&=&\sum\limits_{i_s=1}^{N_x^s}\sum\limits_{j_s=1}^{N_y^s}\frac{\sqrt{\gamma}_\idx{i,j}}{N_x^s N_y^s} max\left(0, \eta_\idx{i,j}^\idx{n}-\tilde{b}_\idx{i,j,i_s,j_s}\right),  \\
	\Ht\left(\eta_\idx{i+\half,j}^\idx{n}\right)&=&\sum\limits_{j_s=1}^{N_y^s}\frac{\sqrt{\gamma}_\idx{i+\half,j}}{N_y^s} max\left(0, \eta_\idx{i+\half,j}^\idx{n}-\tilde{b}_\idx{i+\half,j,j_s}\right),  \\
	\Ht\left(\eta_\idx{i,j+\half}^\idx{n}\right)&=&\sum\limits_{i_s=1}^{N_x^s}\frac{\sqrt{\gamma}_\idx{i,j+\half}}{N_y^s} max\left(0, \eta_\idx{i,j+\half}^\idx{n}-\tilde{b}_\idx{i,j+\half,i_s}\right). 
	\label{eq:variable_dual_2}
\end{eqnarray}
Of course the Jacobian of $\Ht\left(\xi\right)$ should be computed accordingly and reads
\begin{eqnarray}
	G^{(l)}_\idx{i,j}=\sum\limits_{i_s=1}^{N_x^s}\sum\limits_{j_s=1}^{N_y^s}\frac{\sqrt{\gamma}_\idx{i,j}}{N_x^s N_y^s} \mathcal{H}\left(\xi_\idx{i,j}^{(l)}>\tilde{b}_\idx{i,j,i_s,j_s}\right)
	\label{eq:Gdisc}
\end{eqnarray}
where $\mathcal{H}(l)$
is the Heaviside function. The structure of the resulting non-linear system is essentially the same, the Newton algorithm is then guaranteed to converge in a finite number of iterations \citep{Casulli2019}, since $N_x^s$ and $N_y^s$ are finite  if the initial guess satisfy $\xi_\idx{i,j}>\max\limits_{i_s, j_s} \tilde{b}_\idx{i,j,i_s,j_s}$ for all $i$ and $j$.

%====================================================================
\subsection{Explicit non linear terms}
\label{sec.Fm}
\begin{figure}[!h]
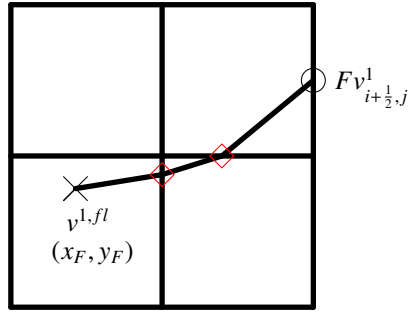

	\centering
	\figB
	\caption{Lagrangian trajectory in the uniform mesh.}
	\label{fig:semiLagrangian}
\end{figure}

We need now to specify the computation of $F^{1,\idx{n}}_{\idx{i+\half,j}}$ and $F^{2,\idx{n}}_{\idx{i,j+\half}}$ for each $i=1\ldots N_x$ and $j=1\ldots N_y$. Concerning the pure convective step, any conservative explicit scheme can be used, such as a simple upwind \citep{Ferrari2021}. However this would lead to a time step restriction based on the local coordinate velocity. %While on Euclidean or cylindrical spaces this may be acceptable, it becomes rather inconvenient in spherical coordinates since, close to the poles, coordinate velocities are huge with respect to the physical ones. 
% Comment (in all our test cases we are ok even with explicit, so it is better not to emphatize it)
In order to obtain an unconditionally stable method, here we use a simple semi-Lagrangian approach following the idea introduced by \cite{Stelling2003}, and more recently by \cite{Tavelli2022} and \cite{CASULLI2026}. Namely  we adopt a simple semi-Lagrangian approach based on a modified advection speed as depicted in Fig. \ref{fig:semiLagrangian}. First we compute a new field $v^{1,*}_\idx{i+\half,j}$ and $v^{2,*}_\idx{i,j+\half}$ using $v^1_\idx{i+\half,j}$, $v^2_\idx{i,j+\half}$ and $\Ht^{n}_\idx{i,j}$, see \cite{Tavelli2022} for details. Then, for every interface $(\idx{i+\half,j})$, we move backward until the foot of the Lagrangian trajectory. The velocity field at the foot, $(v^{1,fl},v^{2,fl})$, is then obtained using a simple interpolation of the surrounding velocities. Then we can simply set
\begin{eqnarray}
	F^{1,\idx{n}}_{\idx{i+\half,j}}=v^{1,fl} \Ht^{n}_\idx{i+\half,j} &\qquad& F^{2,\idx{n}}_{\idx{i,j+\half}}=v^{2,fl} \Ht^{n}_\idx{i,j+\half}\,.
	\label{eq:41}
\end{eqnarray}
The metric correction from the right hand side of Eq.~\eqref{eq:CSW-2-f}
is added to \eqref{eq:41} in the form  of an explicit contribution: 
\begin{eqnarray}
	F^{1,\idx{n}}_{\idx{i+\half,j}}&=&v^{1,fl} \Ht^{n}_\idx{i+\half,j}+ \frac{\Delta t}{2} \gamma^{11}_\idx{i+\half,j}\left[\left(v^{1,\idx{n}}_\idx{i+\half,j}\right)^2 (\partial_1 \gamma_{11})(x_\idx{i+\half},y_\idx{j})+\left(v^{2,\idx{n}}_\idx{i+\half,j}\right)^2 (\partial_1 \gamma_{22})(x_\idx{i+\half},y_\idx{j})\right] \Ht^{n}_\idx{i+\half,j} \\
	F^{2,\idx{n}}_{\idx{i,j+\half}}&=&v^{2,fl} \Ht^{n}_\idx{i,j+\half}+ \frac{\Delta t}{2} \gamma^{22}_\idx{i,j+\half}\left[\left(v^{1,\idx{n}}_\idx{i,j+\half}\right)^2 (\partial_2 \gamma_{11})(x_\idx{i},y_\idx{j+\half})+\left(v^{2,\idx{n}}_\idx{i,j+\half}\right)^2 (\partial_2 \gamma_{22})(x_\idx{i},y_\idx{j+\half})\right] \Ht^{n}_\idx{i,j+\half}.
\end{eqnarray}
The Coriolis force can then be added by solving the advective-Coriolis subsystem \eqref{Coriolis-1a}-\eqref{Coriolis-2b} or better, by considering also the metric correction terms as an additional explicit part of the subsystem \eqref{Coriolis-1a}-\eqref{Coriolis-2b}.
%==================================================================
\subsection{Scaling for large scale simulations}
\label{sec.Scaling}
When this covariant approach is  applied  to the full earth,
a proper rescaling of the equation becomes necessary, since 
the quantities in Eq.~\eqref{eq:metric_spherical} are in the order of $10^{13}$
in the standard units of meters.
% and seconds. 
In order to see how the rescaling works,
let first explicit the factor $\sqrt{\gamma}$ in the PDE system \eqref{eq:CSW-1}-\eqref{eq:CSW-2}:
\begin{eqnarray}
	\label{eq:Tube_1a}
	&&\partial_t (\sqrt{\gamma} H) +\partial_\beta (\sqrt{\gamma} H v^\beta)=0\,, \\
	\label{eq:Tube_1b}
	&&\partial_t (\sqrt{\gamma} H v_\alpha)+\partial_\beta \left(\sqrt{\gamma} H v_\alpha v^\beta \right)+g\sqrt{\gamma} H \partial_\alpha \eta=\frac{1}{2}\sqrt{\gamma} H v^\mu v^\nu\partial_\alpha \gamma_{\mu\nu}\,.
	%-\frac{1}{2}\sqrt{\gamma} H^2 \partial_\alpha g\,.
\end{eqnarray}
In spherical coordinates, $\sqrt{\gamma}=R^2\sin\theta$, hence it is convenient to 
divide both the continuity and the momentum equation by the reference radius $R^2$. The new system reads
\begin{eqnarray}
	\label{eq:Tube_1a_r}
	&&\partial_t (\sqrt{\tilde{\gamma}} H) +\partial_\beta (\sqrt{\tilde{\gamma}} H v^\beta)=0\,, \\
	\label{eq:Tube_1b_r}
	&&\partial_t (\sqrt{\tilde{\gamma}} H v_\alpha)+\partial_\beta \left(\sqrt{\tilde{\gamma}} H v_\alpha v^\beta \right)+g\sqrt{\tilde{\gamma}} H \partial_\alpha \eta=\frac{1}{2}\sqrt{\tilde{\gamma}} H v^\mu v^\nu\partial_\alpha \gamma_{\mu\nu}\,,
	%-\frac{1}{2}\sqrt{\gamma} H^2 \partial_\alpha g\,.
\end{eqnarray}
where 
\begin{equation}
	\tilde{\gamma}_{\alpha\beta }(\theta,\phi)=\frac{1}{R^2}
	\begin{pmatrix}
		R^{2} & 0\\[6pt]
		0        & R^{2}\sin^{2}\theta
	\end{pmatrix}
	=\begin{pmatrix}
	1 & 0\\[6pt]
	0        & \sin^{2}\theta
	\end{pmatrix}
	\label{eq:metric_spherical_rescaled}
\end{equation}
is a rescaled metric tensor.  
It is worth mentioning that this practical approach amounts to the replacement  $\sqrt{\gamma}\rightarrow \sqrt{\tilde{\gamma}}$,
while the transformation among covariant and contravariant components is 
still performed using the true metric given by Eq.~\eqref{eq:metric_spherical}.
Several practical tests have shown that this pragmatic procedure is quite efficient,
and the Newton algorithm is able to handle the equations with a reasonable tolerance, namely $tol_N=10^{-8}$, for all the considered test cases.
%===================================================================
\subsection{Algorithm adaptation to arterial flows}
\label{sec.Tube}
\begin{figure}[!h]
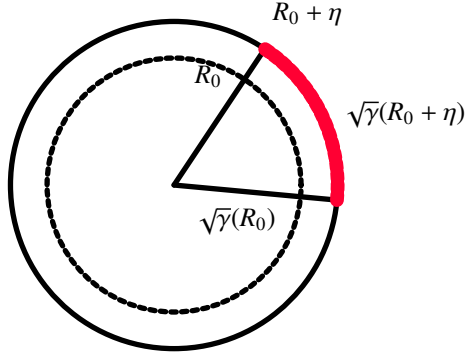

	\centering
	\figA
	\caption{Draft of the cylindrical geometry and metric}
	\label{fig:Tube1}
\end{figure}

It this section we want to show that a classical semi-implicit scheme for arteries can be obtained directly from the proposed algorithm using cylindrical coordinates and an appropriate re-interpretation of the gravity force.
Let us start from the PDE system written with the explicit contribution of the metric as in Eq. \eqref{eq:Tube_1a}-\eqref{eq:Tube_1b}
and with the  metric in standard cylindrical coordinates:
\begin{equation}
	\gamma_{\alpha\beta}= \left(\begin{array}{cc}
		1 & 0 \\
		0 & R^2
	\end{array}\right) \qquad
	\sqrt{\gamma}=R\,,\qquad x^1=z, x^2=\theta\,.
\end{equation}
It is clear that if we want to represent an elastic pipe, expansions or contractions of the pipe will be modeled  
through a change of the metric along the radial direction, which, in our two-dimensional framework, corresponds to the passive direction, see Fig. \ref{fig:Tube1}. For convenience, we use the same definitions introduced by \cite{CasulliArtery2012} and \cite{Lucca2023} for arteries. 
The pressure is related to the radius through the expression $p=p_e+\beta(R-R_0)$ where $\beta$ is a rigidity coefficient  \citep{CasulliArtery2012}. In \cite{Lucca2023} this relation was expressed in terms of an area as
\begin{eqnarray}
	\label{eq:arterial_p}
	p&=&p_e+ k \left(\sqrt{\frac{A}{A_0}}-1\right) = p_e +\frac{k}{\sqrt{A_0}}(\sqrt{A}-\sqrt{A_0})=p_e+\frac{k}{R_0}(R-R_0)=p_e+\beta(R-R_0),
\end{eqnarray}
where we have defined $\beta=k/R_0$. If we combine Eq.~\eqref{eq:arterial_p} with the 
hydrostatic approximation adopted throughout this paper, i.e.  $p-p_e=\rho g\eta$, we obtain
\begin{eqnarray}
	R=R_0+\frac{1}{\beta}(p-p_e)=R_0+g\rho_0 \frac{\eta}{\beta}=R_0+\eta\,,
\end{eqnarray}
where we have re-defined $g=\beta/\rho_0$. The net effect of this approach is that the metric acquires a dependence on $\eta$ which expresses the deformation of the tube:
\begin{equation}
	\gamma_{\alpha\beta}(z,\theta,R_0,\eta)  = \left(\begin{array}{cc}
		1 & 0 \\
		0 & (R_0+\eta)^2
	\end{array}\right) \qquad
	\sqrt{\gamma}=R_0+\eta=R
	\qquad \frac{\partial\sqrt{\gamma}}{\partial \eta} =1\,.
	\label{eq:metric_tube}
\end{equation}
As usual, we compute $H(\eta)=\max(0, \eta-b)$ and we can set $b=-R_0$. In this way
the bottom collapses to the inner center, and the fluid flow fills the entire cylinder,
which is prone to deformations. 
In addition, $H(\eta)=\max(0,\eta+R_0)=\max(0,R)$. 
Since the metric is no longer constant, we use a simple trapezoidal rule to approximate the metric term along the radial direction, namely
\begin{equation}
%	\int_{-R_0}^\eta\sqrt{\gamma}\,dz\approx  \frac{1}{2}\left(\left.\sqrt{\gamma}\right|_{\eta=-R_0}+\left.\sqrt{\gamma}\right|_{\eta})\right)=\frac{R}{2}\,.
	\sqrt{\gamma}\approx  \frac{1}{2}\left(\left.\sqrt{\gamma}\right|_{\eta=-R_0}+\left.\sqrt{\gamma}\right|_{\eta}\right)=\frac{R}{2}\,.
\end{equation}
Neglecting the advection and gravity gradient contributions, Eq.~\eqref{eq:Tube_1a}-\eqref{eq:Tube_1b} become:
\begin{eqnarray}
	&&\partial_t \left(\frac{R^2}{2}\right) +\partial_z \left(\frac{R^2}{2} v^z\right)+\partial_\theta \left(\frac{R^2}{2} v^\theta\right)=0\,, \\
	&&\partial_t\left( \frac{R^2}{2} v_z \right)
	%+\partial_j \left(\sqrt{\gamma} H v_i v^j \right)
	=-g\frac{R^2}{2} \partial_z \eta
	%-\frac{1}{4}R^3 \partial_i g\,. 
 \\
	&&\partial_t\left( \frac{R^2}{2} v_\theta \right)
	%+\partial_j \left(\sqrt{\gamma} H v_i v^j \right)
	= 	-g\frac{R^2}{2} \partial_\theta \eta\,.
	%-\frac{1}{4}R^3 \partial_i g\,.
	\label{eq:Tube_2} 
\end{eqnarray}
Integrating the first equation for $\theta \in [0,2\pi]$ and $z=[z_{i-\half},z_{i+\half}]$ and assuming periodic boundaries we obtain
\begin{eqnarray}
	&&\Delta z \, \partial_t (\pi R^2) +\left(\pi R^2 v^z\right)_{z=z_{i+\half}}-\left(\pi R^2 v^z\right)_{z=z_{i-\half}}+\Delta z \left(\frac{R^2}{2} v^\theta\right)_{\theta=2\pi}-\Delta z \left(\frac{R^2}{2} v^\theta\right)_{\theta=0}=0 \nonumber \\
	&&\Longrightarrow \partial_t (\pi R^2) +\frac{\left(\pi R^2 v^z\right)_{z=z_{i+\half}}-\left(\pi R^2 v^z\right)_{z=z_{i-\half}}}{\Delta z}=0\,,
\end{eqnarray}
which is a consistent integral form of the mass conservation expressed as 
\begin{eqnarray}
	\partial_t A + \pi \partial_z (R^2 v^z)=0\,.
\end{eqnarray}
This shows that we can obtain the one-dimensional elastic tube as a particular case when 
$g=\beta/\rho_0$,
 taking the metric tensor \eqref{eq:metric_tube} with $N_\theta=1$ and $\Delta \theta=2\pi$. Note that, since now the quantity $\Ht(\eta)=\sqrt{\gamma}(\eta)\max(0,\eta-b)$ is in the form of \eqref{eq:Ggammaeta}, then its gradient \eqref{eq:Gdisc} becomes, with pixel resolution,
\begin{eqnarray}
	G^{(l)}_\idx{i,j}=\sum\limits_{i_s=1}^{N_x^s}\sum\limits_{j_s=1}^{N_y^s}\left[
	\frac{\sqrt{\gamma}\left(\xi_\idx{i,j}^{(l)}\right)}{N_x^s N_y^s} \mathbf{1}\left(\xi_\idx{i,j}^{(l)}>\tilde{b}_\idx{i,j,i_s,j_s}\right)+\frac{1}{N_x^s N_y^s} \left(\frac{\partial\sqrt{\gamma}}{\partial \eta}\right)\left(\xi_\idx{i,j}^{(l)}\right)\max\left(0,\xi_\idx{i,j}^{(l)}-\tilde{b}_\idx{i,j,i_s,j_s}\right)  \right]\,.
	\label{eq:Gdisc2}
\end{eqnarray}
For a proper comparison with a classical solver like in \cite{Lucca2023}, we need to compute the pressure $p^{n+1}-p_e=g\rho \eta^{n+1}=\beta \eta^{n+1}$.
%Even if we can take $N_y=1$ for which $\Delta y=2\pi$ and the discrete equation becomes a consistent finite volume-finite difference approximation of the continuity and momentum equation with elastic pipes, we are not restricted to take $N_y=1$. 
We stress that, by allowing for $N_\theta>1$,
the algorithm is not restricted to axially symmetric flows but it is able to handle non-axisymmetric configurations,
such as those that are produced when the coefficient $\beta$ (and hence $g$) has a dependence on the angle $\theta$. 
\section{Numerical tests}
\label{sec.tests}
\subsection{Smooth wave propagation over waterland}
\begin{figure}[!htbp]
	\begin{center}
		\begin{tabular}{cc} 
			\includegraphics[width=0.49\textwidth]{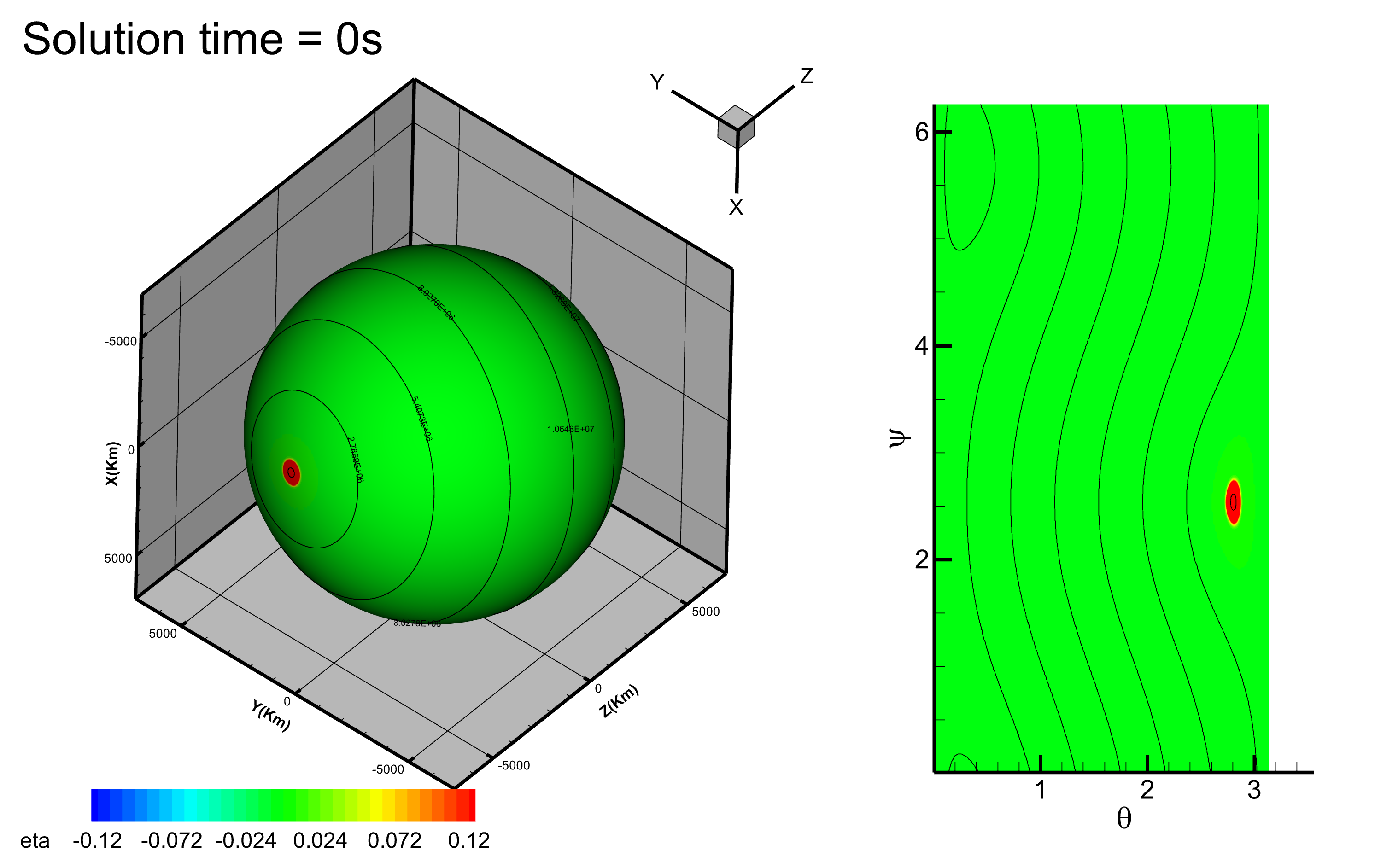}  
			\includegraphics[width=0.49\textwidth]{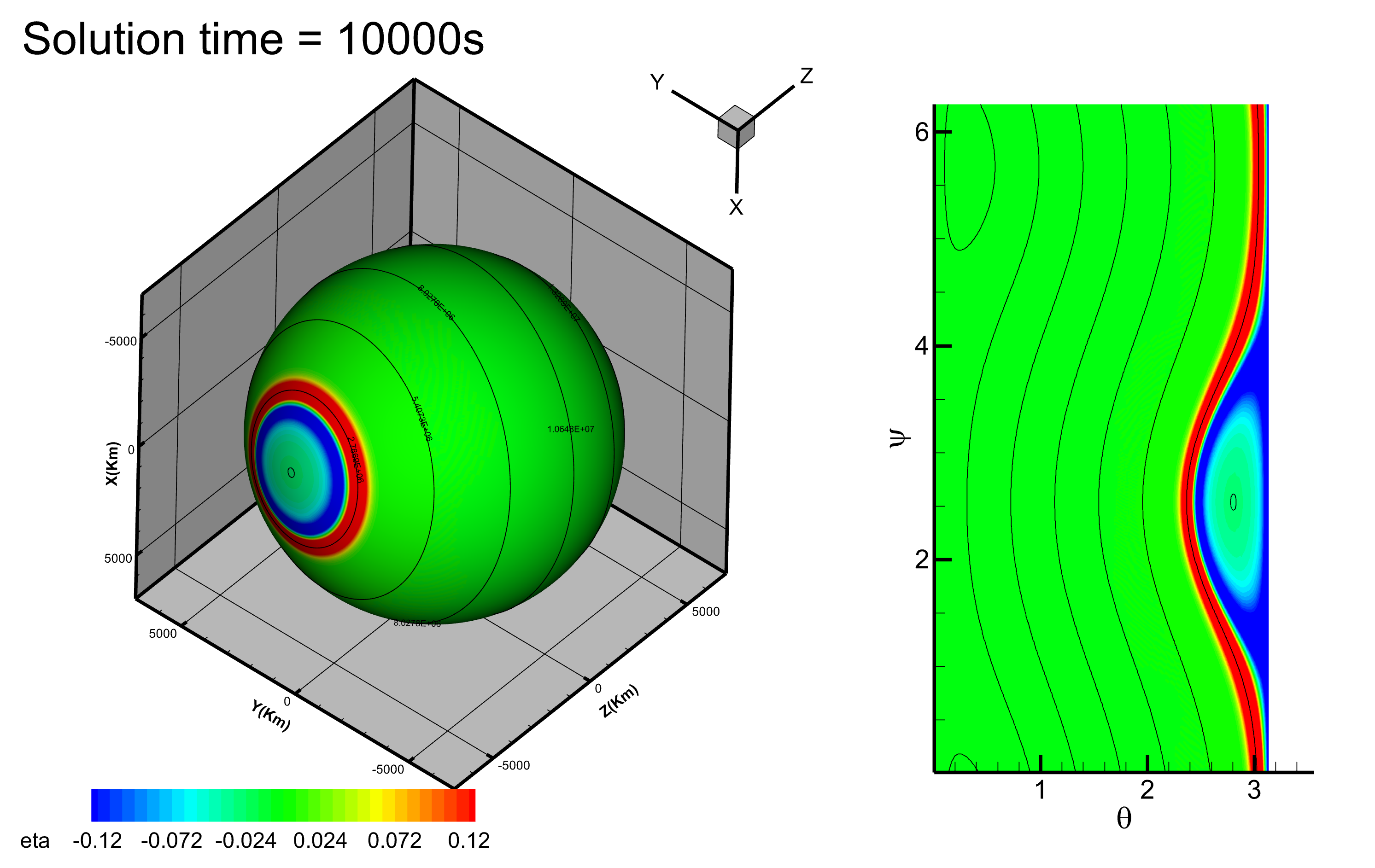}  \\
			\includegraphics[width=0.49\textwidth]{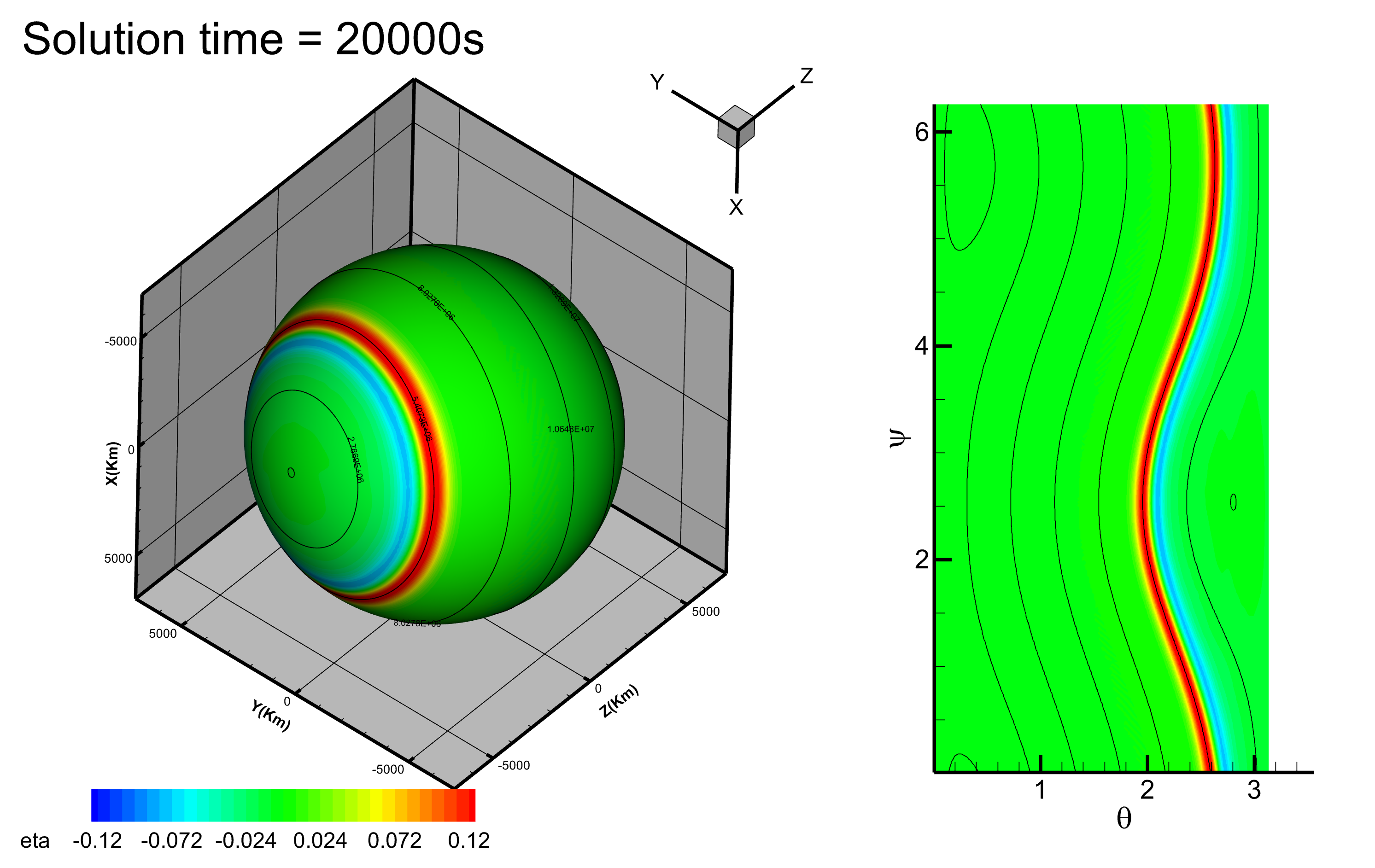}  
			\includegraphics[width=0.49\textwidth]{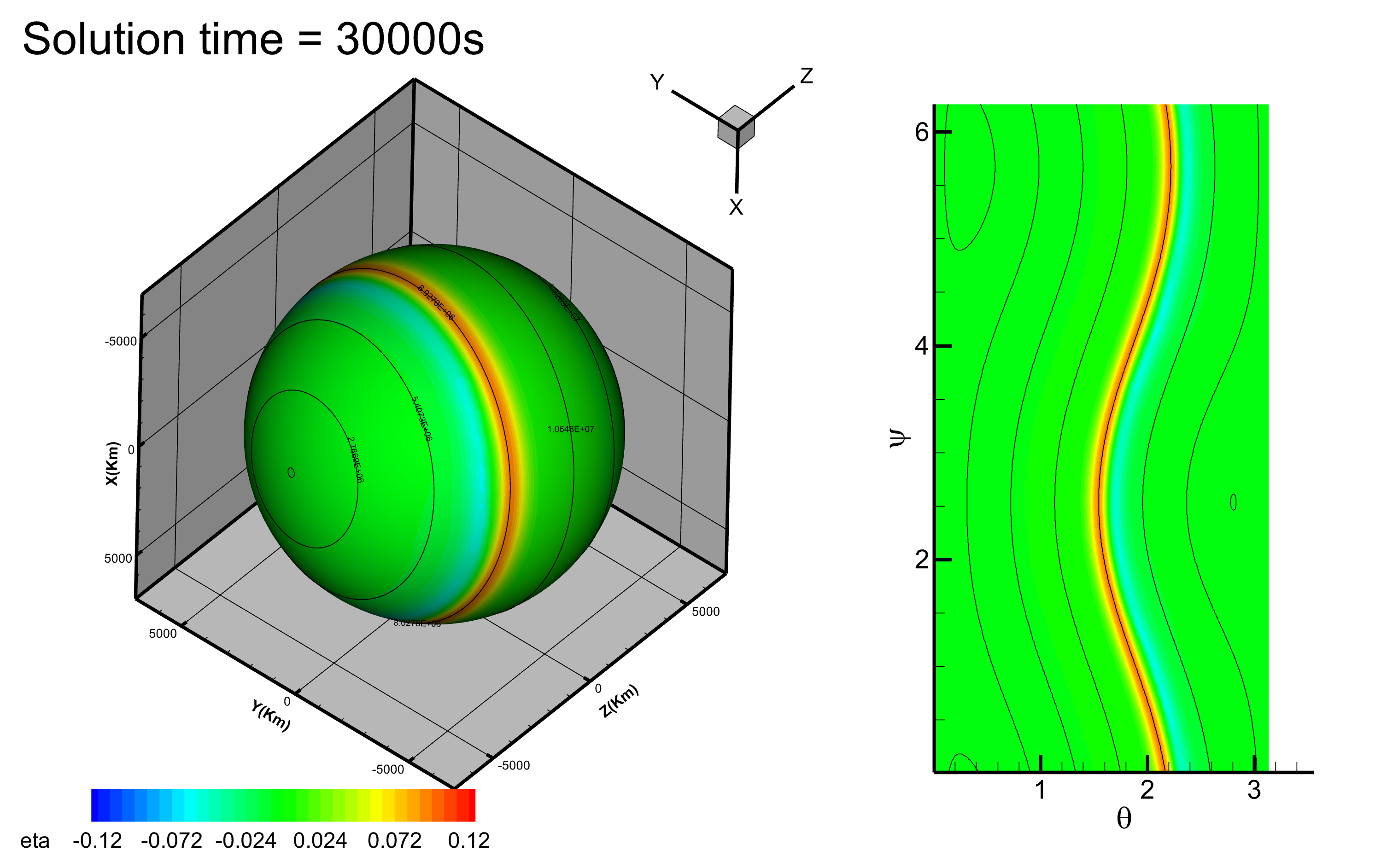}  \\
			\includegraphics[width=0.49\textwidth]{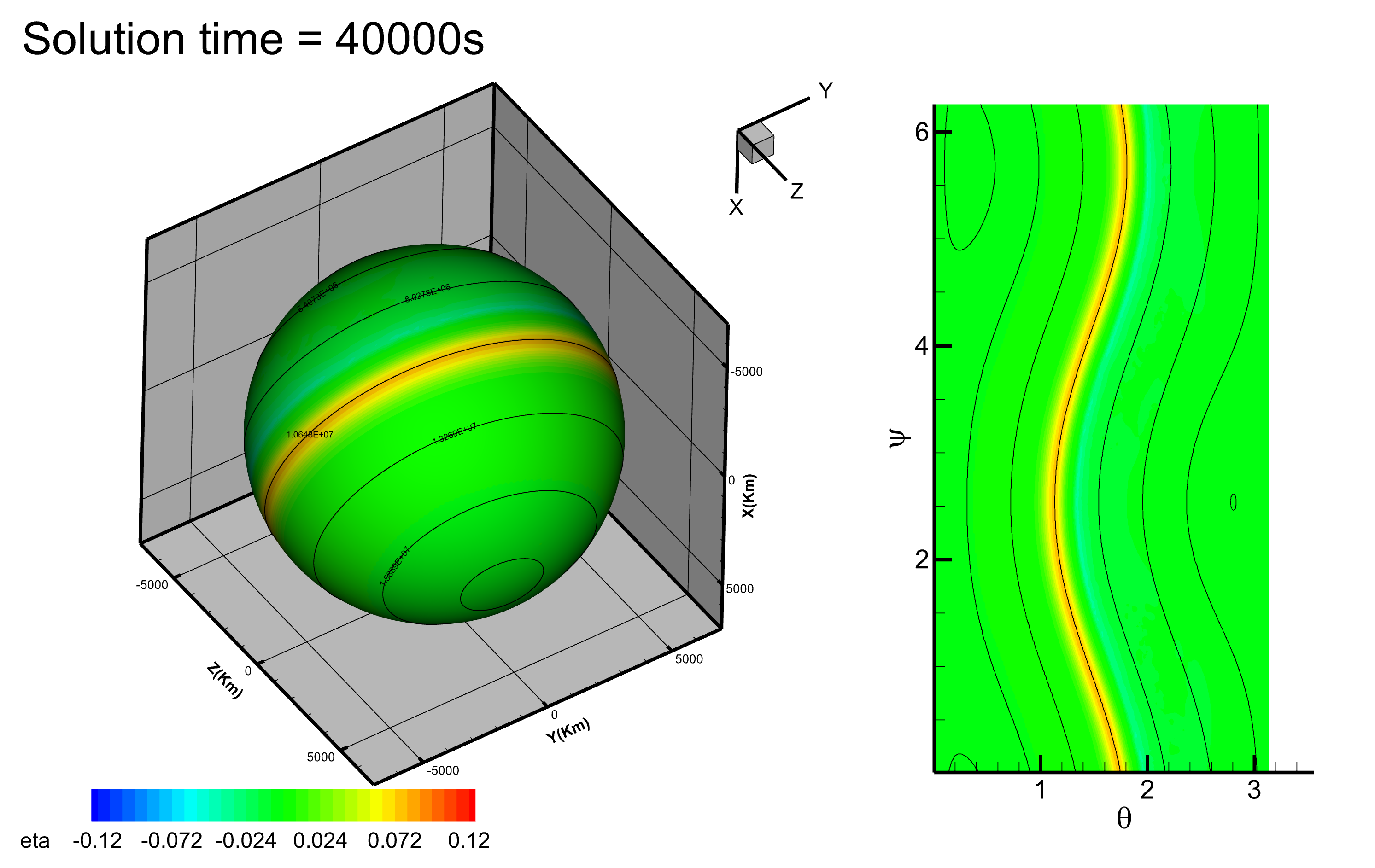}  
			\includegraphics[width=0.49\textwidth]{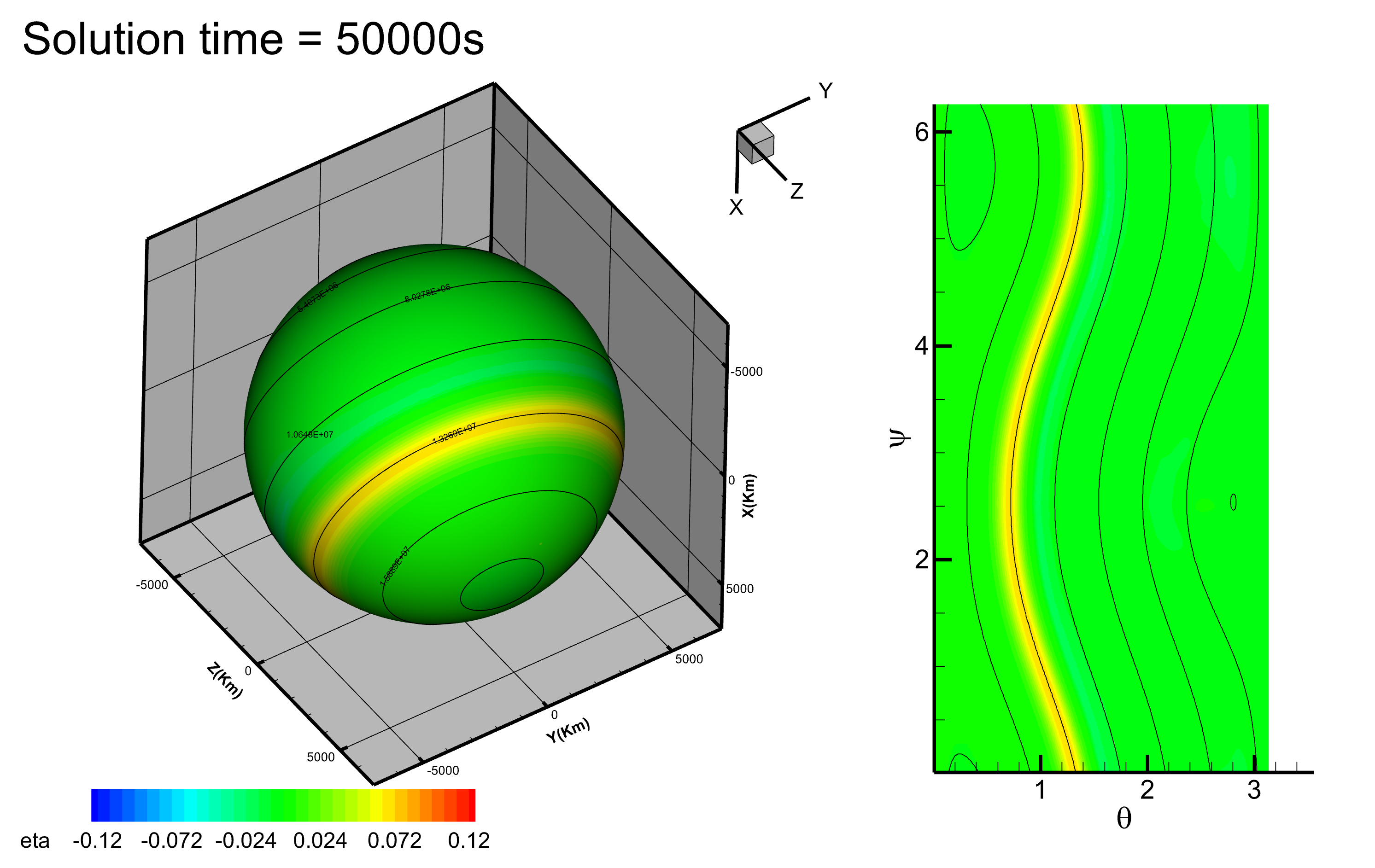} \\
			\includegraphics[width=0.49\textwidth]{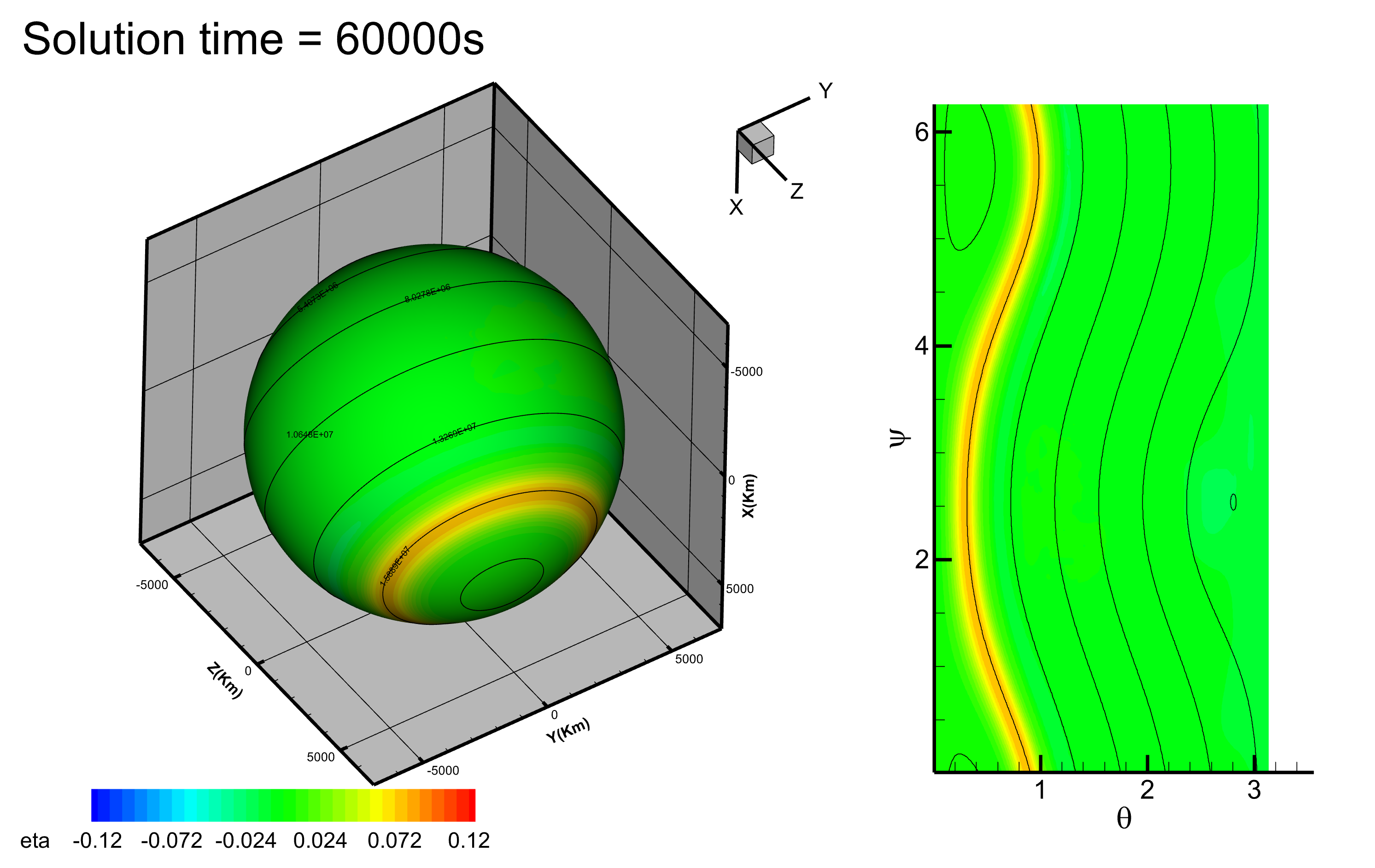}  
			\includegraphics[width=0.49\textwidth]{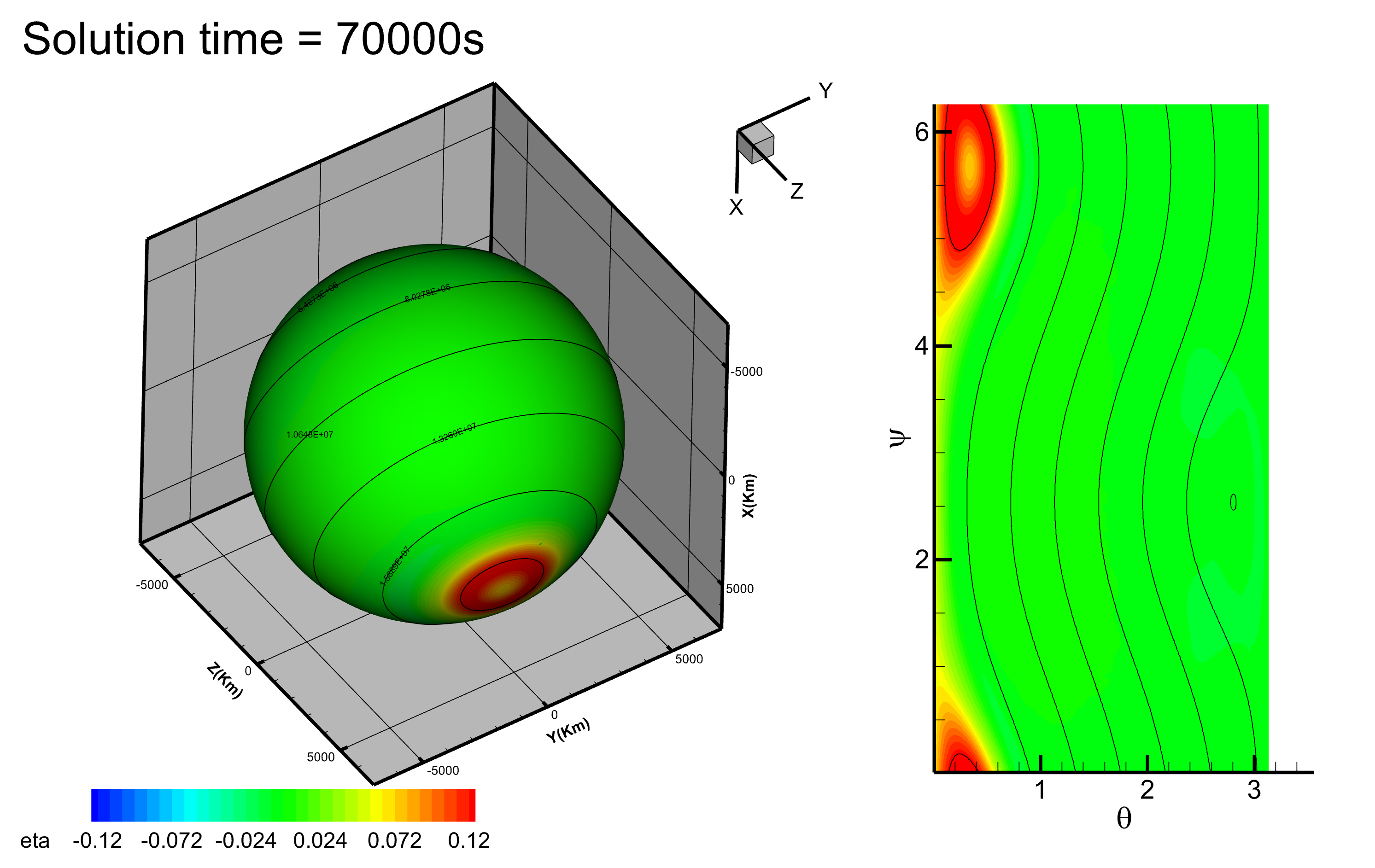}  
		\end{tabular} 
		\caption{Time evolution of the wave at times $t=[0,1\times10^4,2\times10^4,3\times10^4,4\times10^4,5\times10^4,6\times10^4,7\times10^4]\textrm{s}$, from top left to bottom right along rows. The free surface elevation is shown both in the coordinate space and on the manifold.} 
		\label{fig.WL1}
	\end{center}
\end{figure}

In this section we want to test a simple smooth wave propagation at very large scales. In particular, we are interested in the behavior of a travelling wave when it crosses the poles, where the metric factor $\sqrt{\gamma}\rightarrow 0$. In this framework, even if the real velocity of the fluid is moderate, the coordinate velocity may be very large to compensate the distortion due to the metric. We assume the metric in our standard spherical coordinates:
\begin{equation} 	
\gamma_{\alpha\beta}(\theta,\phi,R)= \left(\begin{array}{cc}
	R^2 & 0 \\
	0 & R^2\sin^2\theta
\end{array}\right)
\label{eq:spherical}
\end{equation}
with $R=6.371\cdot 10^6$m and the total heigh is assumed negligible with respect to this manifold, i.e. $\tilde{H}<<R$. To do so, the bathymetry is taken constant on the entire sphere and it is equal to $b=-7000m$. For this test, Coriolis force is deactivated. 
The initial conditions are given by  a simple Gaussian profile centered in $\theta_0 = 160^\circ 00' 00''$ and $\phi_0=-35^\circ 00' 00''$ which corresponds to the standard latitude and longitude given by $70^\circ \mathrm{S}$ and $35^\circ \mathrm{W}$. The shape of the free surface is then defined by
\begin{equation}
	\eta(\theta,\phi)=10 e^{-\ell^2(\theta,\phi,\theta_0,\phi_0)/\sigma^2}\,,
\end{equation}
with $\sigma=2 \cdot 10^5 m$, where $\ell(\theta,\phi,\theta_0,\phi_0)$ is the geodetic distance among  two points on the surface defined by the coordinates $(\theta,\phi)$ and $(\theta_0,\phi_0)$.

The coordinate space is then defined by $\Omega_R=[0,\pi]\times[0,2\pi]$ and it is covered by $N_x=200$ and $N_y=400$ elements. For the time discretization we use
$\Delta t=100s$.
The resulting celerity is $c=\sqrt{gH}\approx 262.049\,\textrm{m/s}$. 
Note that an explicit discretization of the celerity would be limited by this remarkably high velocity.
We consider $t_{end}=8\times 10^4\textrm{s}$ with a sampling output time given by $\Delta t_{out}=10^4\textrm{s}\approx 2 \,\,\textrm{hours}\,\, 46\,\, \textrm{min}$. 
In Figure \ref{fig.WL1} is reported the wave evolution in both the coordinate space and on the manifold, with snapshots at different times displayed along the rows in the figure.
We  also show the isolines of the geodetic distance from the original point $(\theta_0,\phi_0)$, that corresponds to the distance $c\cdot t_{out}$ meters from the source for each output time $t_{out}$.
\begin{figure}[!htbp]
	\begin{center}
		\begin{tabular}{cc} 
			\includegraphics[width=0.49\textwidth]{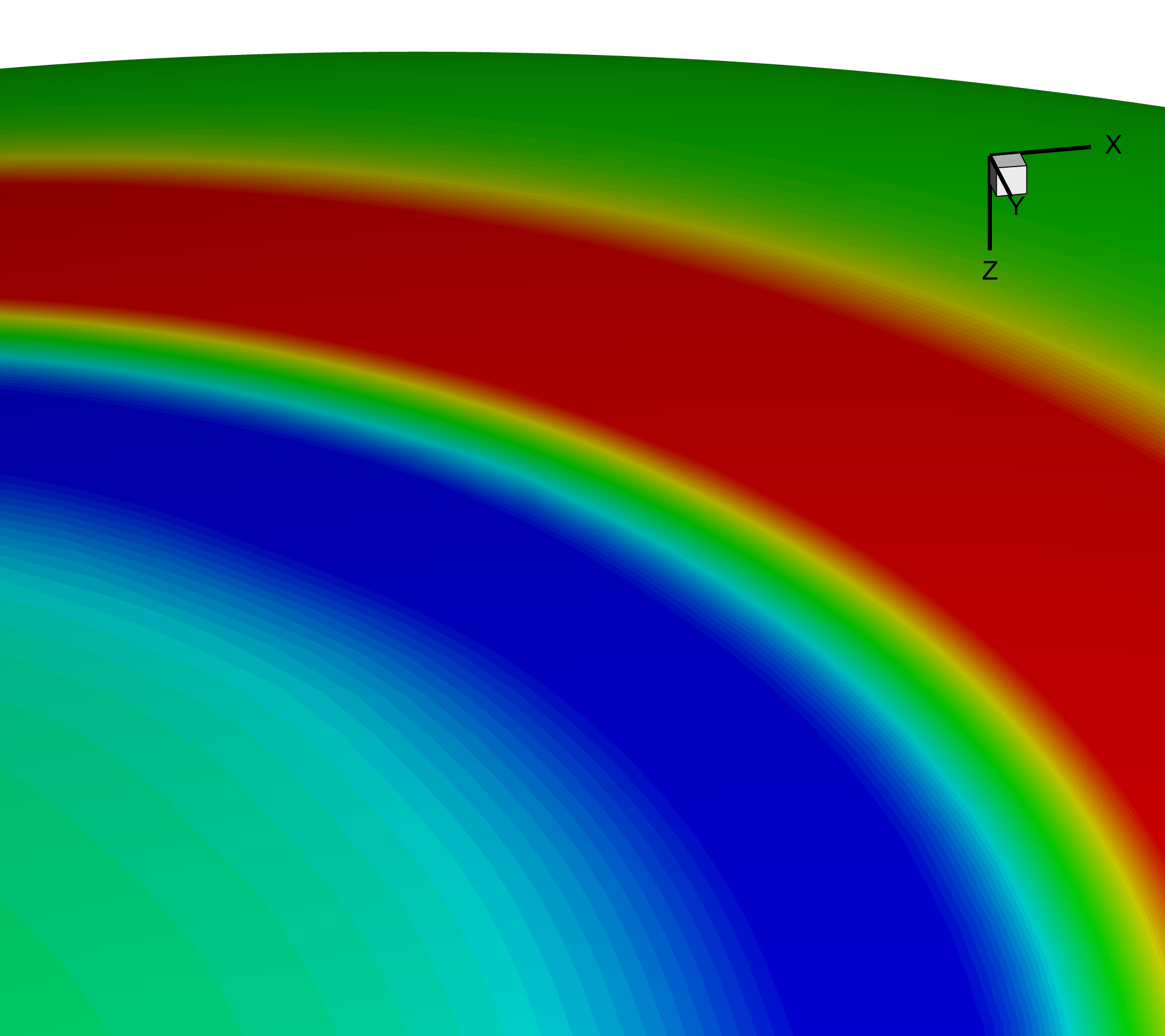}
			\includegraphics[width=0.49\textwidth]{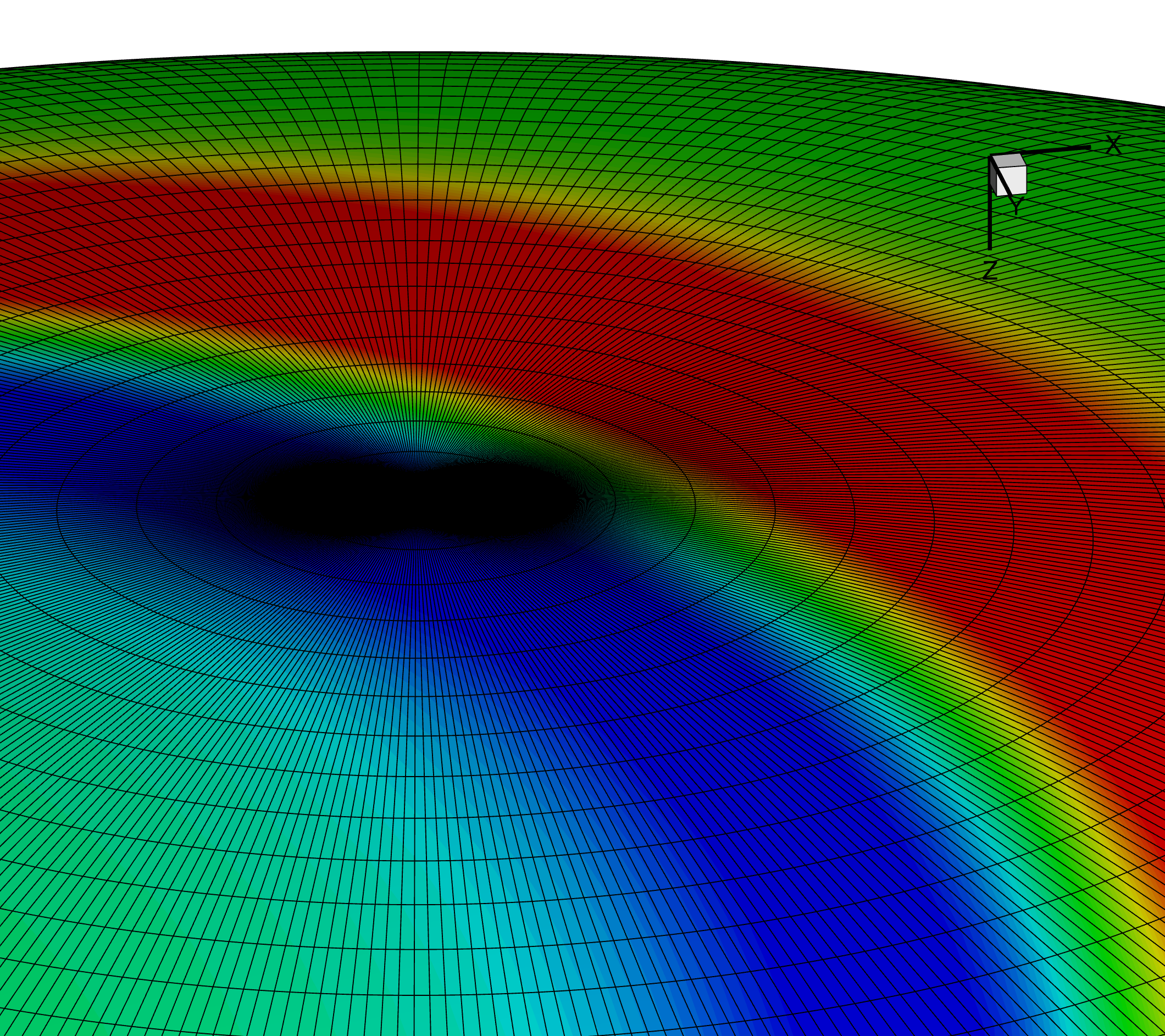}  
		\end{tabular} 
		\caption{Detail of the numerical solution at the south pole at $t=10^4\textrm{s}$ with (right panel) and without the mesh (left panel). The mesh is obtained by connecting the barycenters of each element in the coordinate space.  } 
		\label{fig.WL2}
	\end{center}
\end{figure}
The circular wavefront matches very well the spatial position where it is expected to be at each output time. Moreover, no pathological behaviors occur at the poles, which are formally singular for the coordinates used, 
with no need to adopt special \emph{ad hoc} treatments. 
This peculiar feature is the result of 
%the combined effects of the semi-Lagrangian approach and of 
the Newton algorithm that is able to implicitly solve the metric singularity. 
%In fact, as explained in Sect.~\ref{sec.Fm}, the semi-Lagrangian scheme avoids that the time step is killed by the vanishing cell size at the poles.
In this scenario an explicit discretization, or the semi-Lagrangian scheme summarized in Sect.~\ref{sec.Fm}, are not limiting the time step size since the local fluid velocity is very small. However this is strictly related to the local conformation close to the pole.
 On the other hand, the Newton algorithm, which is a strategy to account for the weak nonlinearity of $H(\eta)\rightarrow 0$, is naturally extended to  $\tilde H(\eta)\rightarrow 0$, which becomes zero at the pole thanks to the metric.
Figure~\ref{fig.WL2} shows the wave front when it crosses the south pole.
Since the mesh is uniform in the coordinate space, it produces an extreme distortion on the manifold, however this does not affect the quality of the wave passing through it. This distortion can also be observed in the reference space (see top-right panel of Figure \ref{fig.WL1}). 

%====================================================================================================
\subsection{Well-balancing}
The method proposed is by construction well-balanced in the sense of the C-property \citep{Bermudez1994,castro2020well,GASSNER2016,Fernández2022} 
as it has been already proved by \cite{Boscheri2023b}.
This property is here verified on two different manifolds.
\begin{enumerate}
	\item
%The first one is 
%a classical test for shallow water model for a water  at rest on a non uniform bathymetry. 
%A well-balanced scheme should be able to preserve the solution $\eta=constant$ and $\vec{v}=0$ exactly. 
We first test the well-balancing property of our numerical method using a classical benchmarks, originally proposed by \cite{leveque1998balancing}. In our version we consider  a domain $\Omega_R=[-0.5,0.25]\times[-0.5,0.5]$ covered with a uniform mesh of size $N_x=400$ and $N_y=200$. 
We obtain their same effective domain by using an ad hoc non-identity metric given by 
\begin{equation}
	\gamma_{\alpha\beta}= \left(\begin{array}{cc}
		16 & 0 \\
		0 & 1
	\end{array}\right)\,,
\end{equation}
so that the resulting physical space becomes 
$\tilde{\Omega}=[-2,1]\times [-0.5,0.5]$. In this way, it is also possible to check the ability of the covariant formulation using a distorted Euclidean space.  The transformation in this case becomes
\begin{eqnarray}
	X(x,y)&=& \sqrt{\gamma_{11}}x  \\
	Y(x,y)&=& \sqrt{\gamma_{22}}y \,.
\end{eqnarray}
 The initial condition is chosen as $v^1=v^2=0$ and
 \begin{equation}
 	\eta(x,y,0)=\left\{
 	\begin{array}{lc}
 		\eta_0+\epsilon & \mbox{if}\,\,\, -0.95 \leq X(x,y) \leq -0.85, \\
 		0 & \hspace{-2.8cm}\mbox{otherwise}
 	\end{array}
 	\right. \qquad b(x,y)=0.8e^{-5(X(x,y)+0.1)^2-50Y(x,y)^2}\,.
 \end{equation}
 We finally use a subgrid resolution of $N_x^s=N_y^s=10$.
 We have run this configuration in two different set up:
 \begin{itemize}
 	\item  $\epsilon=0.0$,  $\eta_0=1$ and $t_{end}=0.1$,  $\Delta t=0.001$
 	\item  $\epsilon=0.01$, $\eta_0=1$ and $t_{end}=0.48$,  $\Delta t=0.001$\,.
 \end{itemize}
In the first setup, where $\eta$ is not perturbed, we directly verify the well-balancing property. The solution at final time is shown in the left panel of  Fig.~\ref{fig.wb1}, while the corresponding errors are reported
in the first line of Tab.~\ref{tab.wellbalancing}. 
In the second set up, with $\epsilon>0$,  we can see the interaction between the traveling wave and the bathymetry as reported in Fig. \ref{fig.wb0} at times $0.12,0.24,0.36$ and $0.48$. One can appreciate that no spurious oscillations are generated during the evolution.
 
 \item
 As a second relevant configuration meant to address well balancing,
 we consider the entire ocean of the earth at rest, using the spherical metric tensor \eqref{eq:spherical} with $R=6.371\cdot 10^6\textrm{m}$. In this case we set $\eta_0=0$, while the bathymetry is built from a real Digital Elevation Model (DEM) based on the data availabe from GebCo \citep{Gebco2025}. 
 In particular, we included the GEBCO 2025 Grid,  with ice surface elevation. The domain in this case is $\Omega_R=[0,\pi]\times [0,2\pi]$, covered with $N_\theta=200$ and $N_\phi=400$ elements while $\Delta t=100\,\textrm{s}$ and $t_{end}=1000\,\textrm{s}$. The corresponding mesh is of $1.57\times 10^{-2} rad$ for the mesh and $1.57\times 10^{-3} rad$ for the pixels, that corresponds to $100\,\textrm{km}$ and $10\,\textrm{km}$ at the equator, respectively.
 The resulting errors in $L^2$ and $L^{\infty}$ are reported in the second line of Tab.~\ref{tab.wellbalancing}. 
 The numerical scheme is able to maintain stationarity up to machine precision, demonstrating the capability of the algorithm to reproduce the well-balancing property. 
 For the spherical case the $L^2$ error is normalized with a sphere of radius $1$, i.e. the element size is computed as $A(\theta,\phi)=\Delta \theta \Delta \phi \sqrt{\gamma(\theta,\phi,1)}$ and the real value should be rescaled with the factor $R^2$. However in this case the $L^\infty$ shows that we are close to machine precision and it is only the real earth surface that generates high values. It is worth mentioning that the adopted tolerances are the same for both cases and the normalization explained in Sect.~\ref{sec.Scaling}, as well as a proper preconditioner, are crucial to ensure convergence of the linear solver and of the Newton algorithm. The right panel of Fig.~\ref{fig.wb1} shows the final steady solution for the \emph{world at rest}.
\begin{table}  
	\caption{$L^2$ and $L^{\infty}$ norm evaluated at $t=t_{end}$ for the Euclidean and the Spherical case.} 	
	\begin{center} 
		\begin{tabular}{ccccc} 
			\hline
			Test & $L^2(\eta)$ & $L^{\infty}(\eta)$ & $L^2(u)$ & $L^{\infty}(u) $\\ 
			\hline
			Euclidean & $1.1843e-14$  & $2.4425e-14$ & $7.3626e-15$ & $2.9043e-14 $ \\
			Spherical &  $1.2180e-12$ & $5.0435e-12$  &  $3.4542e-14$ & $3.1491e-13$ \\
			\hline           
		\end{tabular}
	\end{center}
	\label{tab.wellbalancing}
\end{table} 

\begin{figure}[!htbp]
	\begin{center}
		\begin{tabular}{cc} 
			\includegraphics[width=0.45\textwidth]{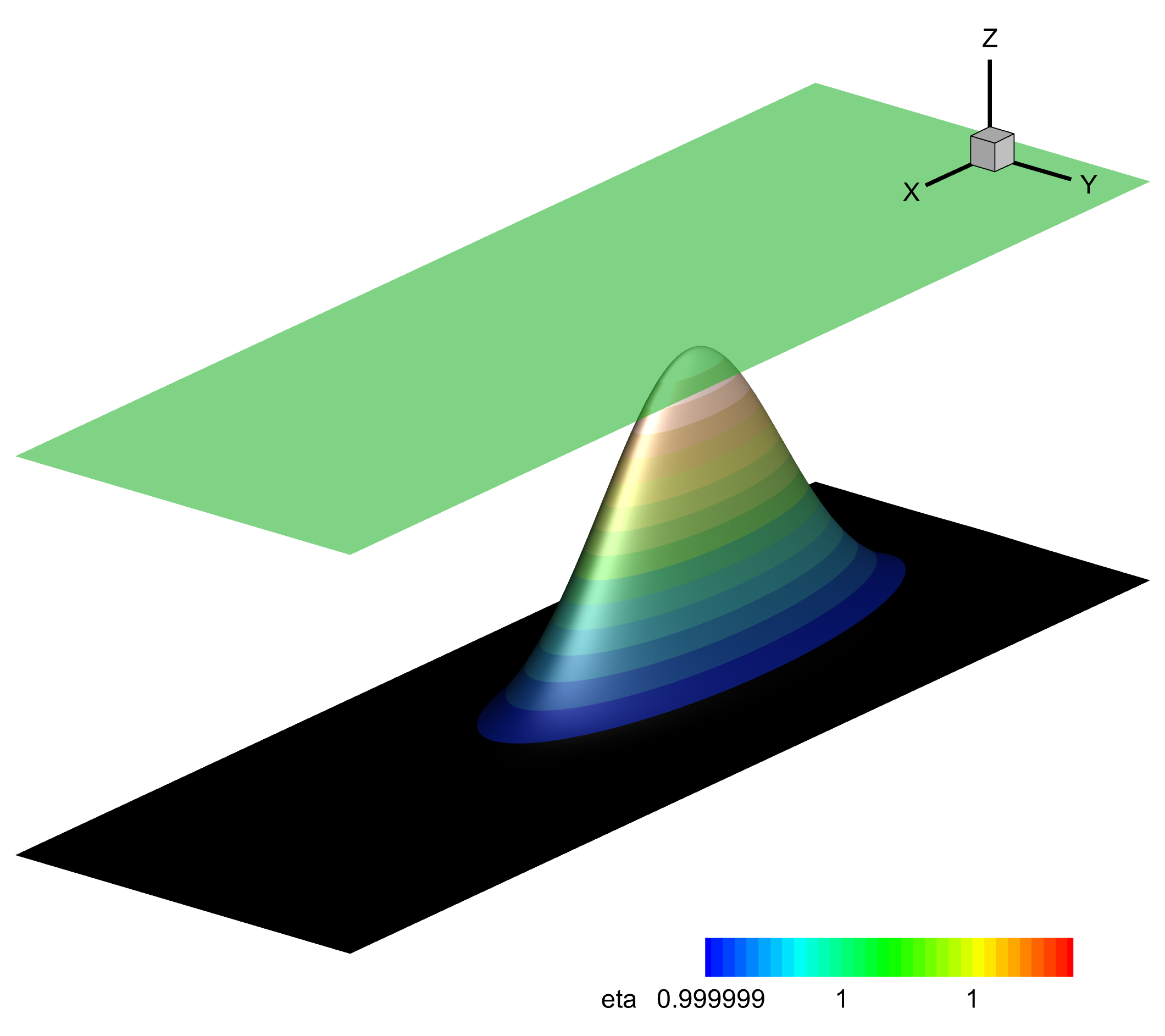}  
			\includegraphics[width=0.45\textwidth]{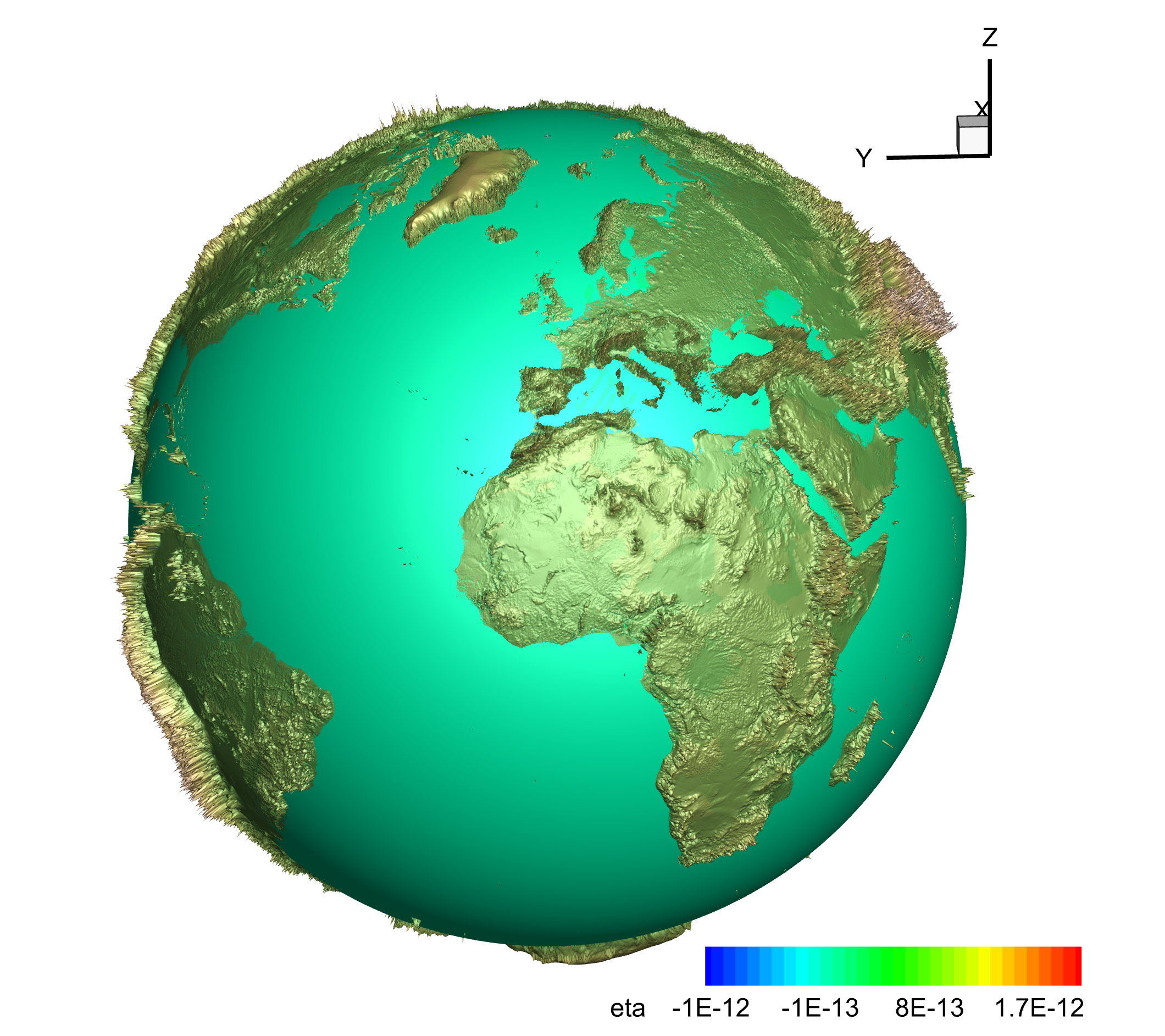} 
		\end{tabular} 
		\caption{Steady equilibrium using Euclidean space at time $t_{end}=0.1$ (left) and Spherical coordinates at $t_{end}=1000$ (right) with real DEM data. In the right plot the bathymetry is rescaled with a factor $100$ for better visualization. } 
		\label{fig.wb1}
	\end{center}
\end{figure}
\begin{figure}[!htbp]
	\begin{center}
		\begin{tabular}{cc} 
			\includegraphics[width=0.45\textwidth]{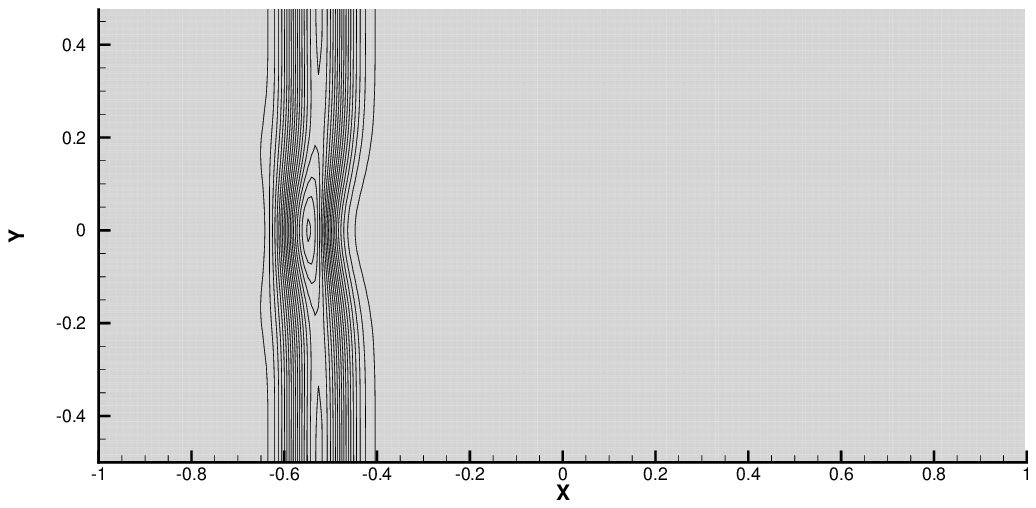}  
			\includegraphics[width=0.45\textwidth]{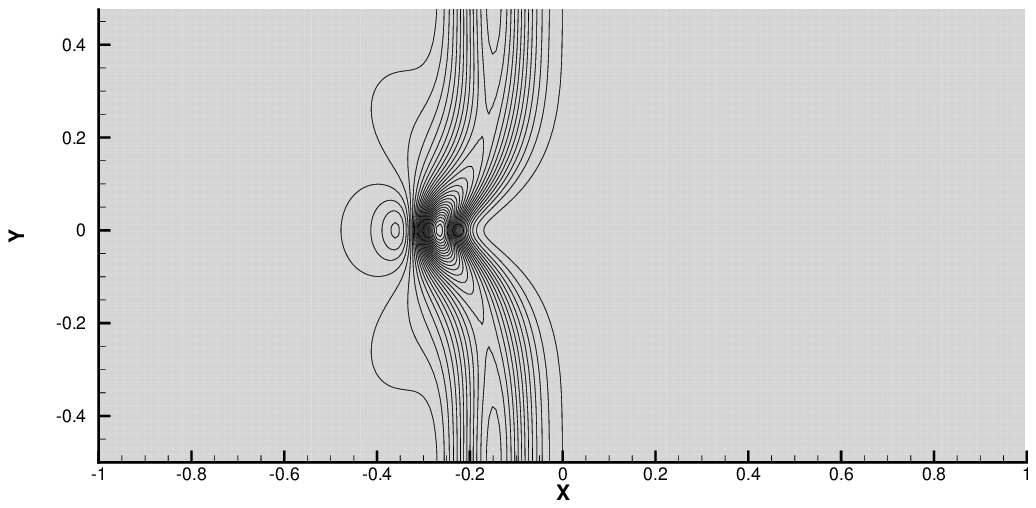} \\
			\includegraphics[width=0.45\textwidth]{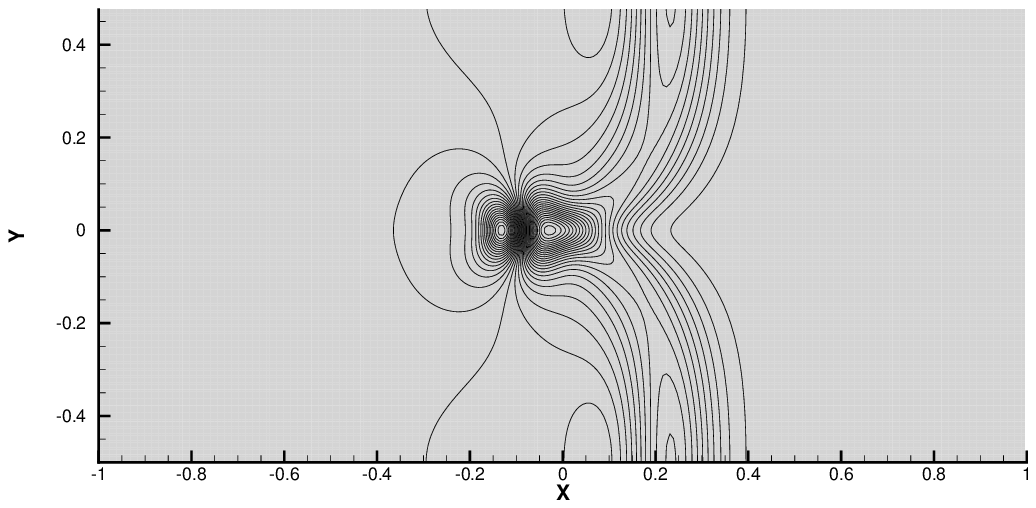}  
			\includegraphics[width=0.45\textwidth]{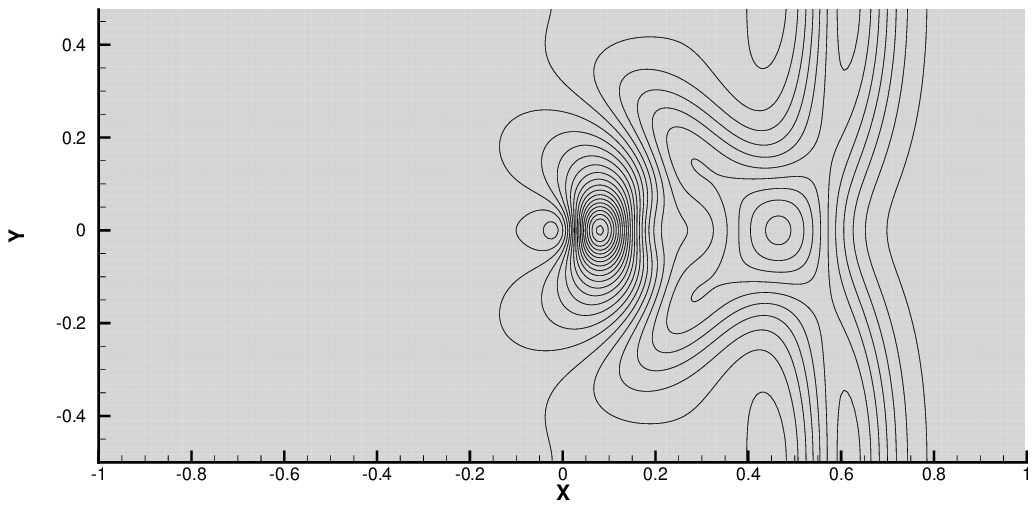}
		\end{tabular} 
		\caption{Numerical solution with $\epsilon=10^{-2}$ at times $t=0.12$, $t=0.24$, $t=0.36$ and $t=0.48$, from top left to bottom right. We report the free surface elevation $\eta$ though $80$ equidistant contour lines ranging in the interval $\eta \in [0.99,1.01]$.} 
		\label{fig.wb0}
	\end{center}
\end{figure}

\end{enumerate}

%===============================================================================
\subsection{Riemann problems}
\begin{figure}[!htbp]
	\begin{center}
		\begin{tabular}{cc} 
			\includegraphics[width=0.45\textwidth]{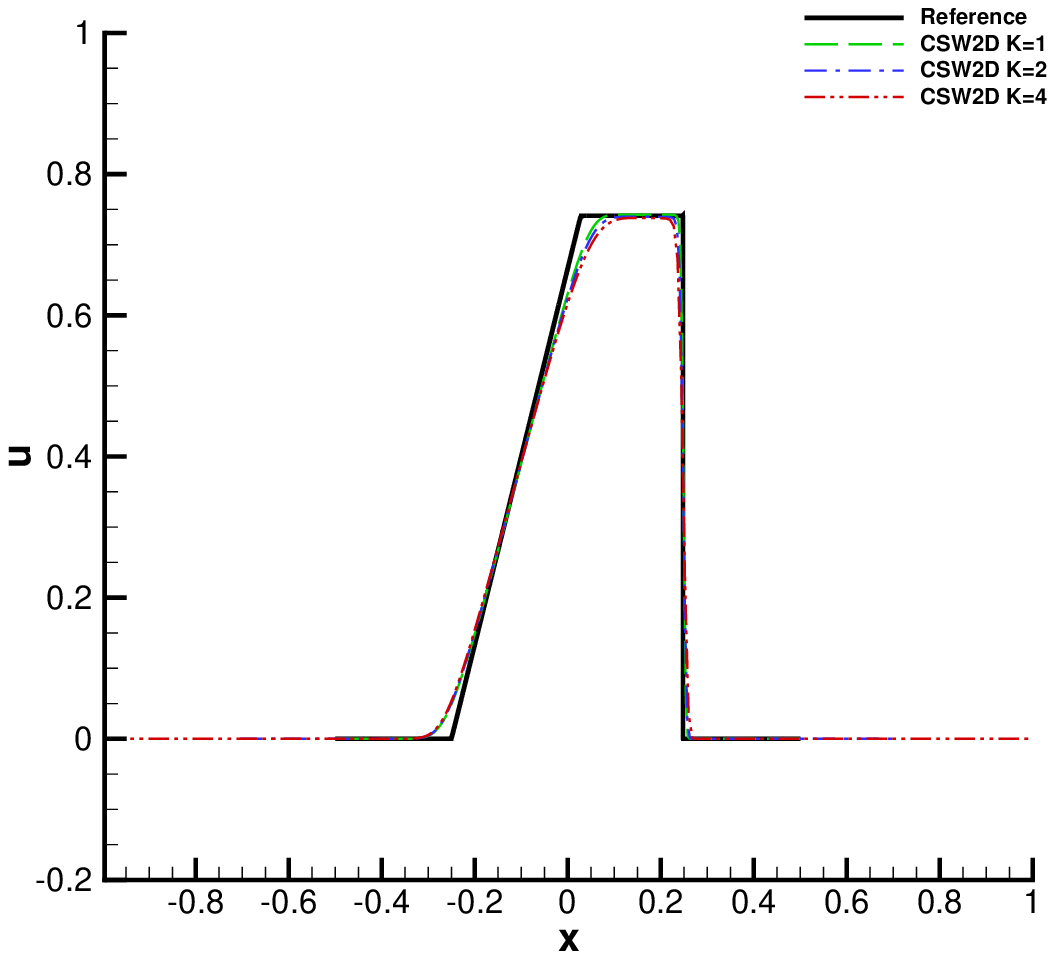}  
			\includegraphics[width=0.45\textwidth]{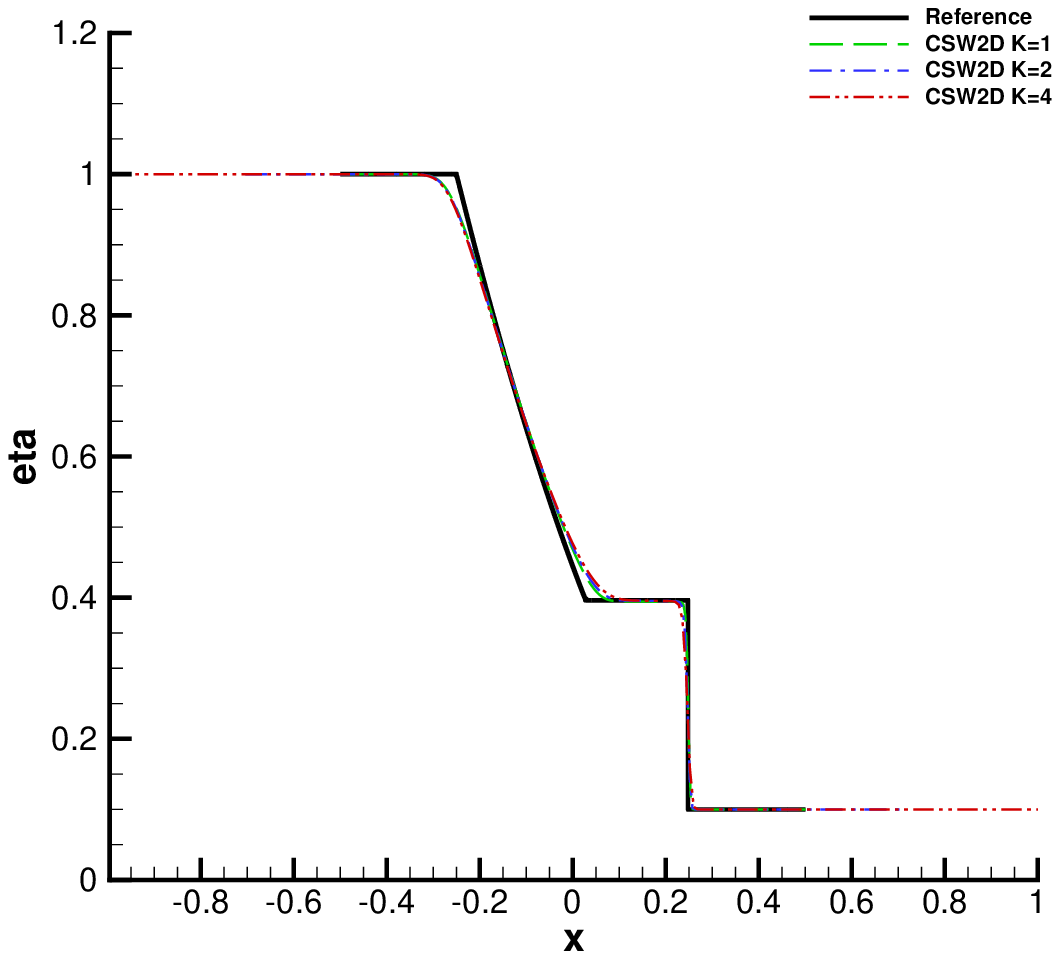}
		\end{tabular} 
		\caption{RP1: Comparison between the numerical and exact solution for the velocity component $u$ (left panel) and free surface $\eta$ (right panel) with different distortion factors $K$.} 
		\label{fig.rp1}
	\end{center}
\end{figure}

\begin{figure}[!htbp]
	\begin{center}
		\begin{tabular}{cc} 
			\includegraphics[width=0.45\textwidth]{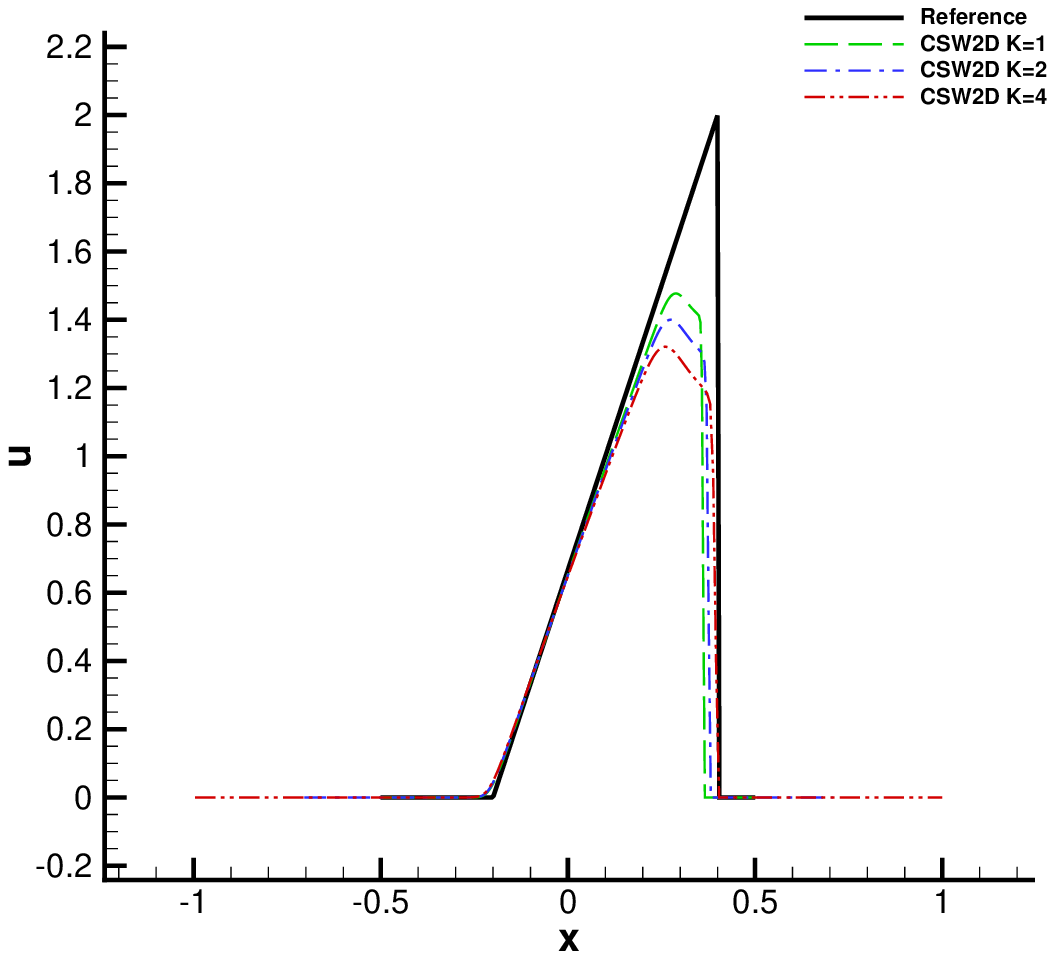}  
			\includegraphics[width=0.45\textwidth]{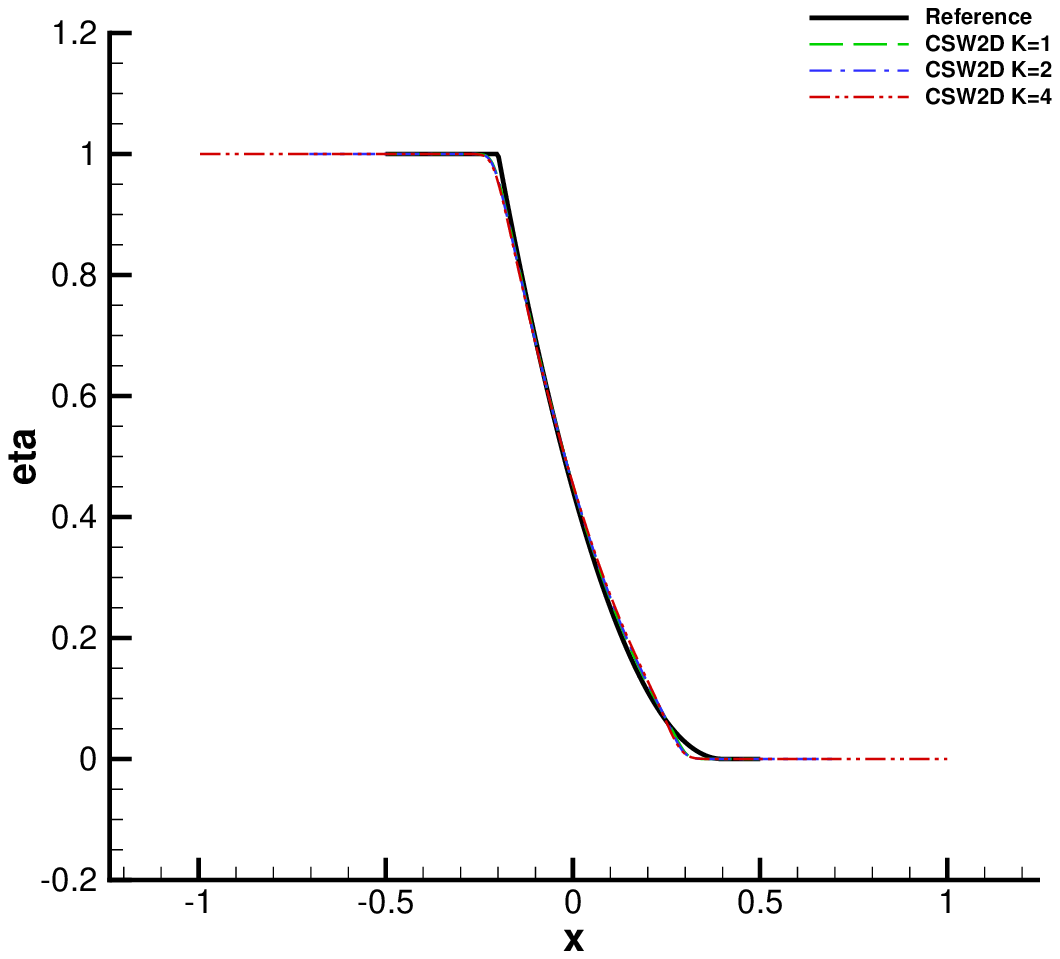}
		\end{tabular} 
		\caption{RP2: Comparison between the numerical and exact solution for the velocity component $u$ (left panel) and free surface $\eta$ (right panel) with different distortion factors $K$.} 
		\label{fig.rp2}
	\end{center}
\end{figure}

In order to verify the capability of the scheme to operate with different metric tensors, we consider a couple of classical Riemann problems \citep{Toro1992,BERNETTI2008,Han2014}: the dam-break over a wet bed and the dam-break over a dry bed. For this test we consider the following Euclidean metric 
\begin{equation}
	\gamma_{\alpha\beta}= \left(\begin{array}{cc}
		K & 0 \\
		0 & 1
	\end{array}\right) \qquad
	\sqrt{\gamma}=\sqrt{K},
\end{equation} 
which is a modification of the classical Euclidean space with a distortion factor $K$ in the $x-$direction. The coordinate domain is set to be $\Omega_R=[-0.5,0.5]\times[0,5]$ which is covered with $N_x=250$ and $N_y=2$ elements. We set $g=1$ while the other parameters can be read in Table \ref{tab.rp}.
\begin{table}  
	\caption{Parameters used in the Riemann problems.} 	
	\begin{center} 
		\begin{tabular}{cccccccc} 
			\hline
			Problem & $\eta_L$ & $\eta_R$   &  $u_L$ & $u_R$ & $b$    & $t_{end}$ & $\Delta t$  \\\hline 
			\hline
			RP1  	&  $1.0$   & $0.1$      & $0.0$  & $0.0$ & $0.0$  & $0.25$    & $0.0025$  \\
			RP2  	&  $1.0$   & $0.0$      & $0.0$  & $0.0$ & $0.0$  & $0.20$    & $0.0010$  \\
			\hline           
		\end{tabular}
	\end{center}
	\label{tab.rp}
\end{table} 
The numerical results obtained for $K=1,2,4$ and the two Riemann problems are reported in Fig.~\ref{fig.rp1} and Fig.~\ref{fig.rp2} against the exact solution taken from \cite{Ferrari2021}.
In all the cases we can see a good agreement with the exact solution, as well as the effect produced by increasing  distortion factors $K$.

%===============================================================================
\subsection{Steady-state nonlinear zonal geostrophic flow}
\begin{figure}[!htbp]
	\begin{center}
		\begin{tabular}{cc} 
			\includegraphics[width=0.45\textwidth]{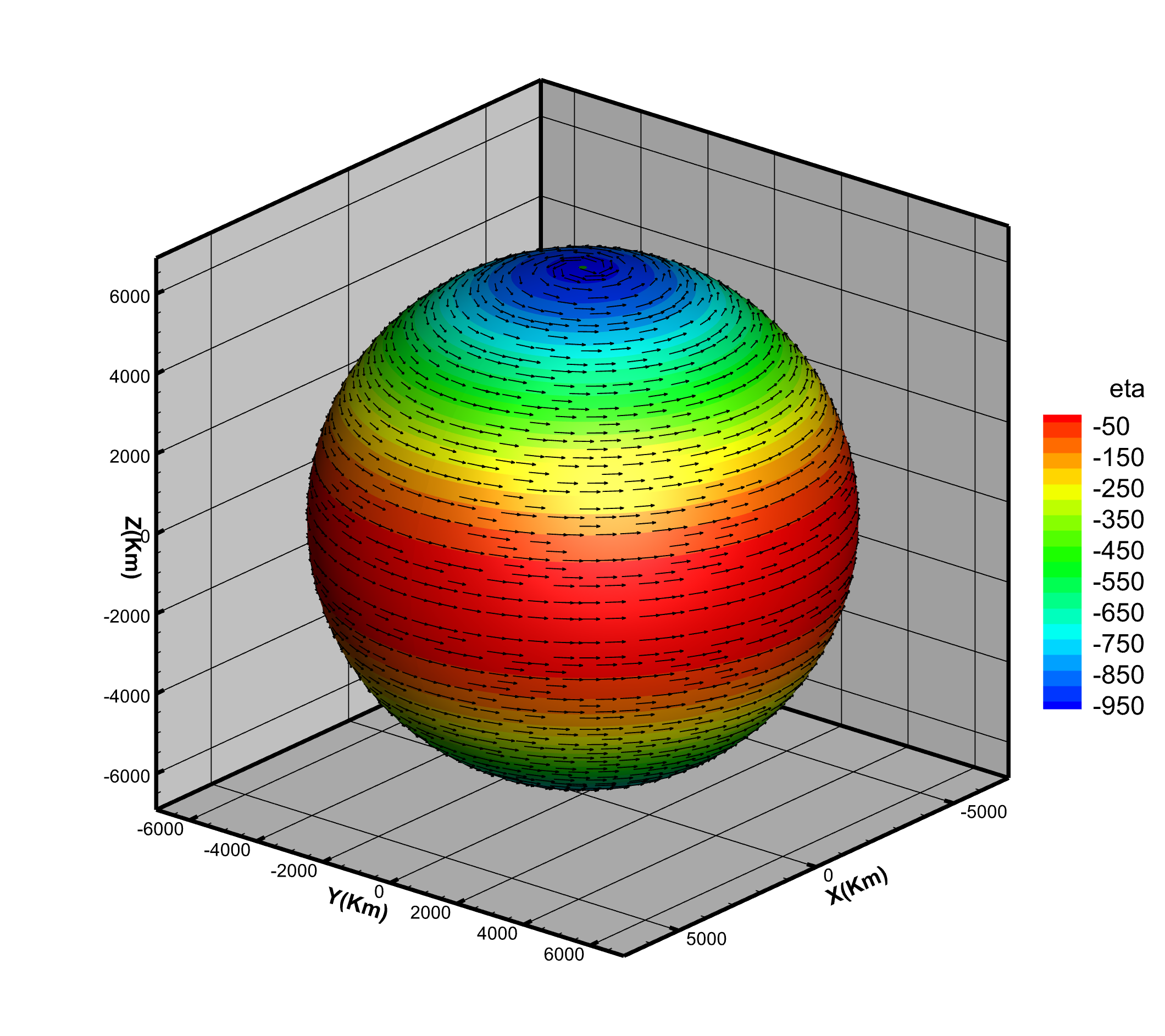}  
			\includegraphics[width=0.49\textwidth]{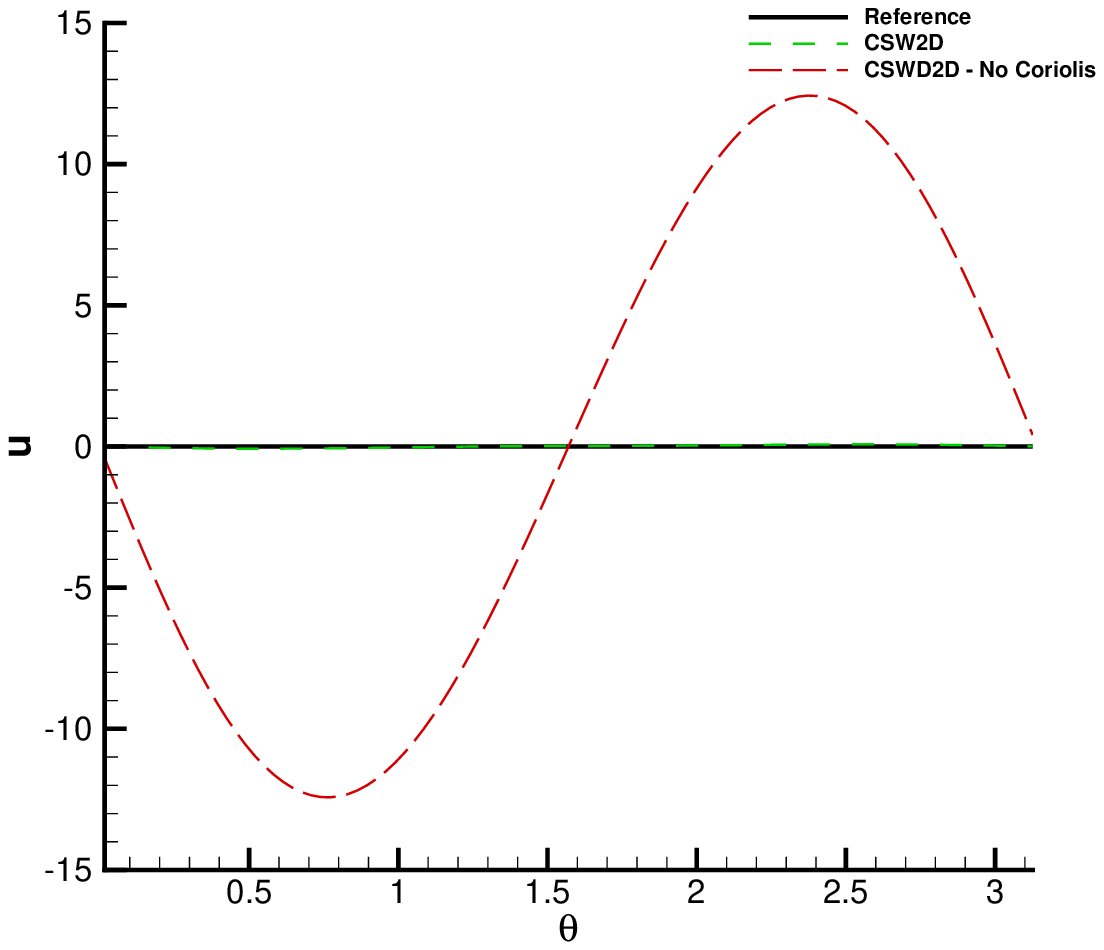} \\
			\includegraphics[width=0.49\textwidth]{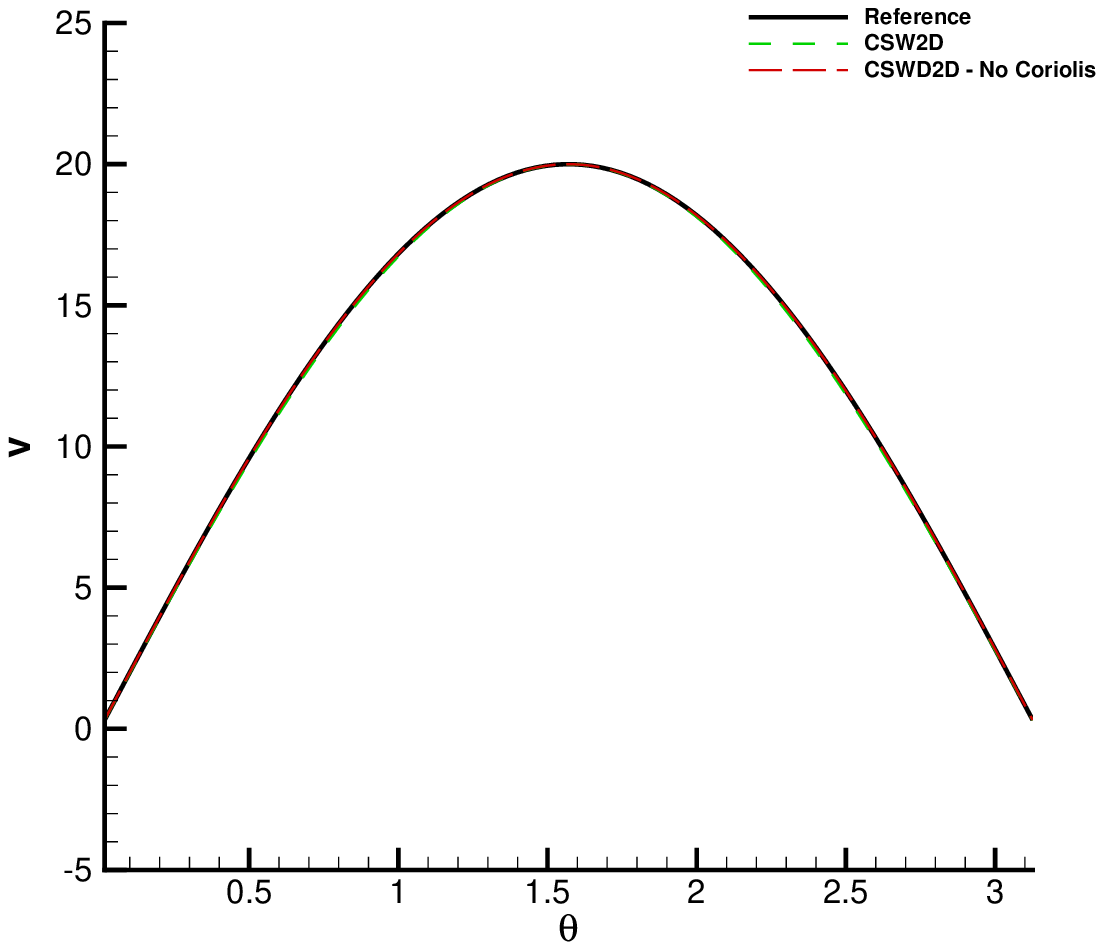}
			\includegraphics[width=0.49\textwidth]{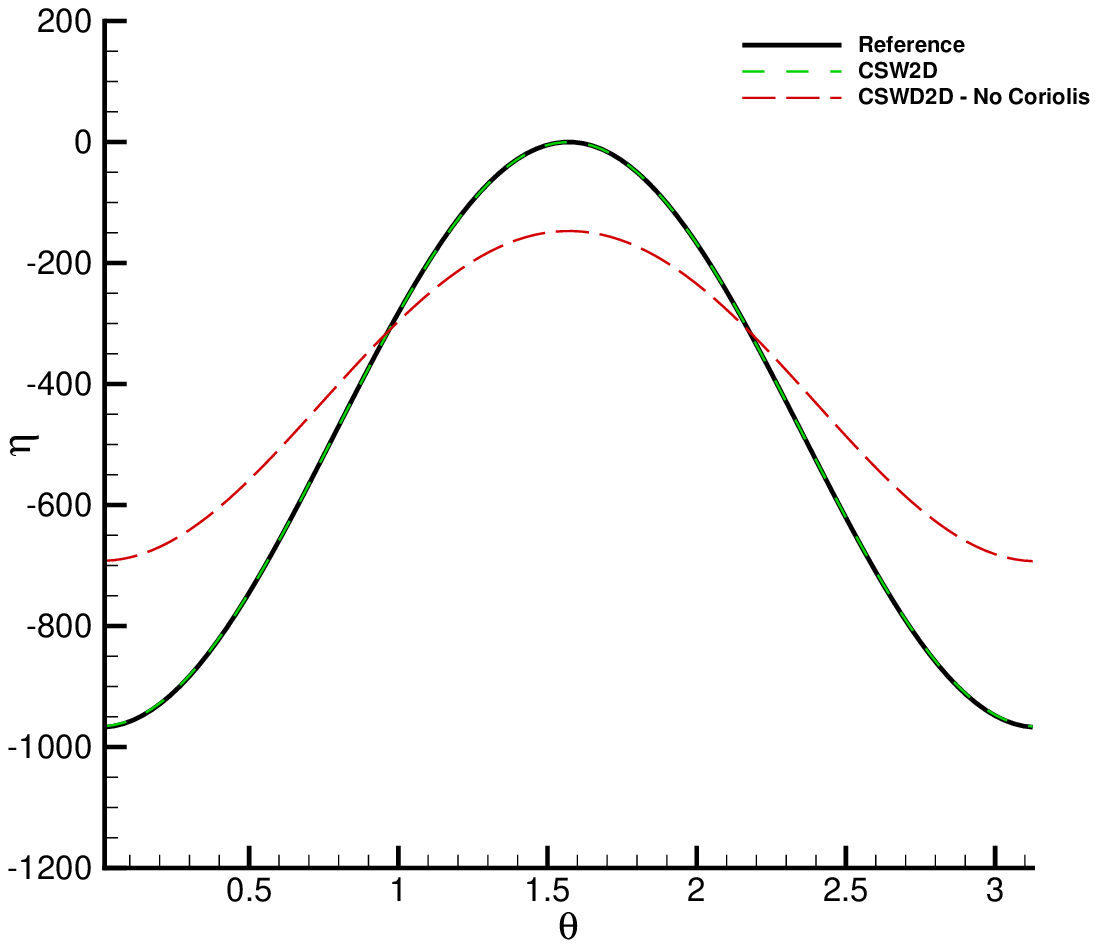}  
			
		\end{tabular} 
		\caption{Top left panel: Free surface elevation and velocity field at $t_{end}=10^4\,s$;  Comparison of the numerical solution with and without the Coriolis effects against the exact steady solution for the velocity field and free surface $\eta$ in the remaining panels.} 
		\label{fig.coriolis2}
	\end{center}
\end{figure}

In order to check the correctness of our semi-implicit scheme on the full equations \eqref{eq:CSW-1-f}-\eqref{eq:CSW-2-f},
including the Coriolis term of Sect.~\ref{sec:Coriolis}, we test our method against an exact solution of the shallow water on a sphere. This test was proposed by \cite{WILLIAMSON1992} and represents a solid body rotation on a sphere where the free surface gradient is in equilibrium with the Coriolis force. The initial condition is set as
\begin{eqnarray}
	u(\theta,\phi,0)&=&0  \\
	v(\theta,\phi,0)&=&U_0 \cos\left(\theta-\frac{\pi}{2}\right)  \\
	\eta(\theta,\phi,0)&=& \eta_0-\frac{1}{g}\left( R\Omega U_0 - \frac{U_0^2}{2}  \right)\sin^2\left(\theta-\frac{\pi}{2}\right) \\
	b(\theta,\phi)&=& b_0.
\end{eqnarray}
For the sphere we use the parameters $R=6.371\times 10^6\,\textrm{m}$, $\Omega_R=2\pi/7.292\cdot 10^{-1}\, \textrm{rad/s}$ and $U_0=20\, \textrm{m/s}$ which refer to the earth parameters. We cover the coordinate domain $\Omega=[0,\pi]\times[0,2\pi]$ with a uniform mesh of size $N_x=100$ and $N_y=200$. We finally set $\Delta t=100\,\textrm{s}$ and $t_{end}=10^4 \,\textrm{s}$. The resulting free surface and velocity field at the final time are reported in Fig.~\ref{fig.coriolis2}, top left panel, for a three dimensional view. The remaining panels of
Fig.~\ref{fig.coriolis2}, on the other hand, allow for a closer comparison to the exact solution,
by showing the velocity field and the free surface elevation, along one dimensional profiles.
In order to quantify the impact of  the Coriolis force, 
we have run the same simulation by removing its presence in the equations. 
It is clear from Fig.~\ref{fig.coriolis2} that the Coriolis force has a huge impact on the entire domain,
and neglecting its contribution leads to un-tolerable errors.

 %===============================================================
\subsection{Pressure gradient over an artery branch}
In this section we want to simulate an artery branch. As a reference we take the RP1 in \cite{Lucca2023}. According to the  notation of Sect.~\ref{sec.Tube}, we use  $k_{ref}=k=2.005\cdot 10^4 \,\textrm{Pa}$, $A_0=3.14\cdot 10^{-4} \textrm{m}^{2}$. The initial pressure is expressed in terms of the  difference between the reference area and the initial area \citep{Lucca2023}. In particular $A_L=f_L A_0$ and $A_R=f_r A_0$. The pressure may be derived as $p_L=k(\sqrt{f_L}-1)$ and $p_R=k(\sqrt{f_R}-1)$ and then $\eta_L=p_L/\beta$, $\eta_R=p_R/\beta$ with $\beta=k/R_0=2.001\cdot 10^6 \,\textrm{Pa/m}$. Setting $\rho_0=1050\, \,\textrm{kg}/\textrm{m}^3$ we get a uniform $g=1.9057\cdot 10^3 \,\textrm{m}/\textrm{s}^2$. The metric tensor is expressed by Eq.~\eqref{eq:metric_tube}
with $b=-R_0$, which is kept constant along the tube. 
A close comparison of our numerical results with those reported in  \cite{Lucca2023} can 
be obtained by simply inverting the previous formula, to get
\begin{eqnarray}
	p=\beta\eta \qquad \frac{A}{A_0}=\frac{\pi (R_0+\eta)^2}{A_0} \qquad f [ml/s]=\pi (\eta+R_0)^2 u\,.
\end{eqnarray}
In our test we use $\Omega_R=[0,0.2]\times[0,2\pi]$ covered with $N_r=400$, $N_\theta=50$. The left and the right axial velocities are taken as in \citep{Lucca2023}: $u_L=1\,\textrm{m/s}$, $u_R=2\,\textrm{m/s}$, while zero angular velocity is considered. The final time is set to $t_{end}=0.013 \,\textrm{s}$ and $\Delta t =10^{-4}\,\textrm{s}$.
% Tecplot commands
%{q[ml/s]}=3.1416*({eta}+0.009997464891747)**2*{ue}*1e6
%{p [KPa]}={eta}*2.001007277006261e+06/1000 
%{A/A0}=({eta}+0.009997464891747)**2/0.009997464891747**2
\begin{figure}[!htbp]
	\begin{center}
		\begin{tabular}{cc} 
			\includegraphics[width=0.45\textwidth]{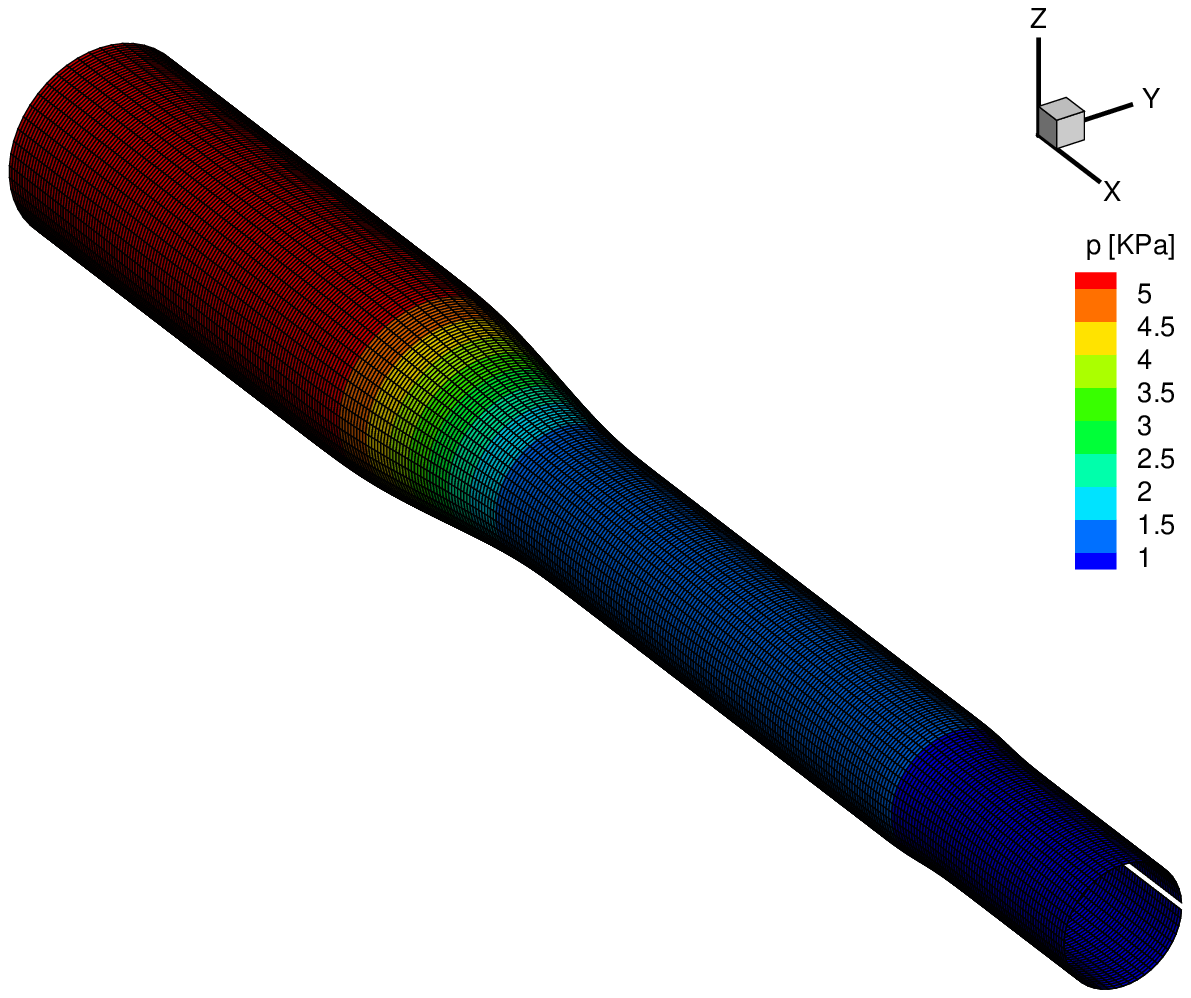}  
			\includegraphics[width=0.45\textwidth]{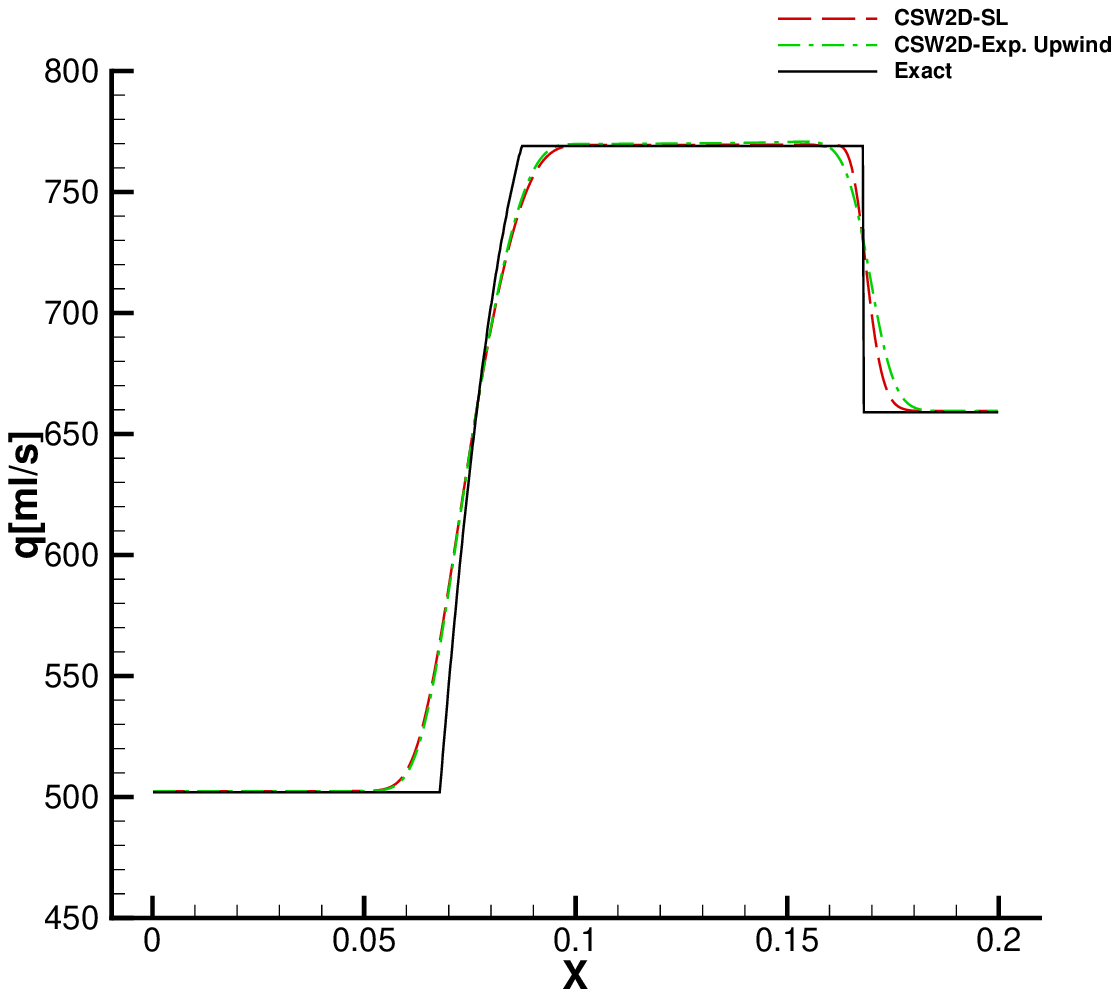} \\
			\includegraphics[width=0.45\textwidth]{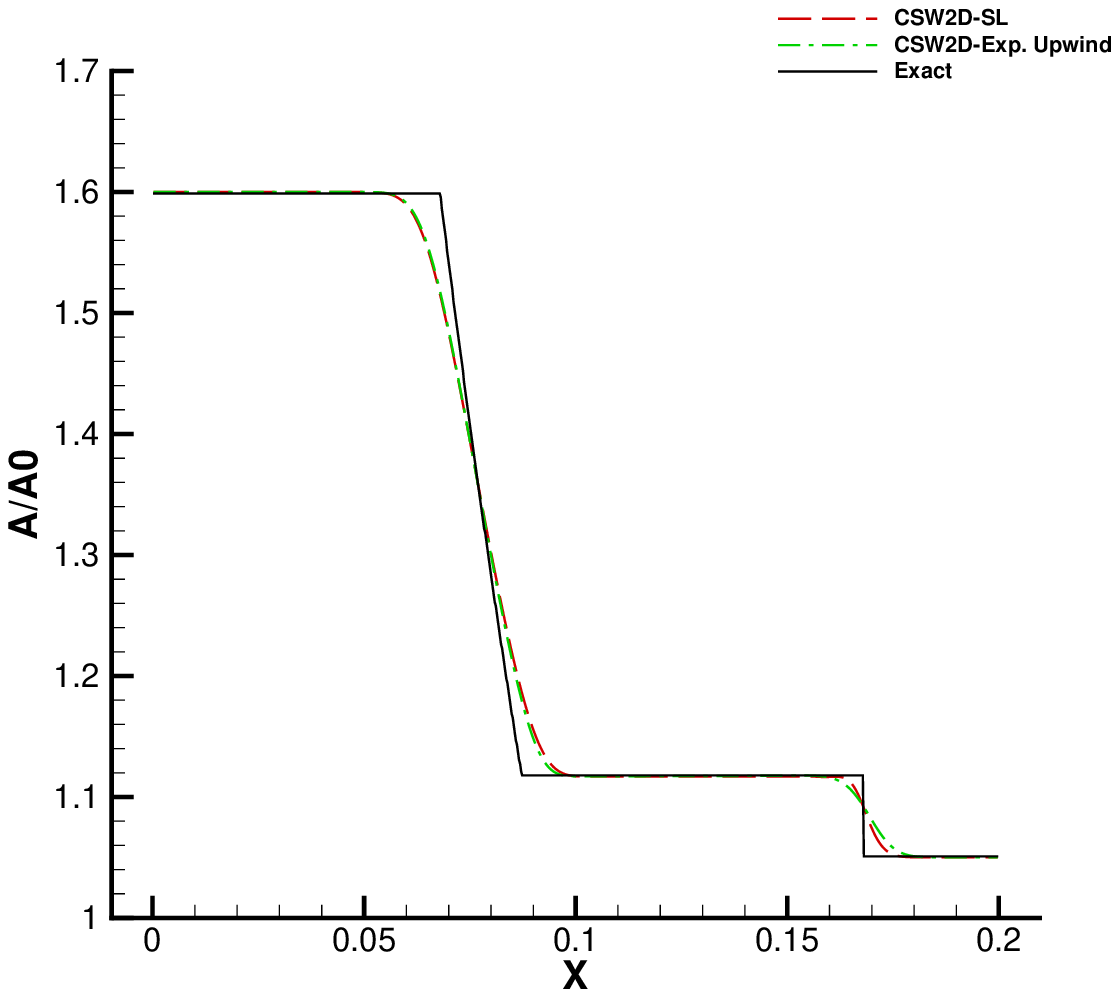}  
			\includegraphics[width=0.45\textwidth]{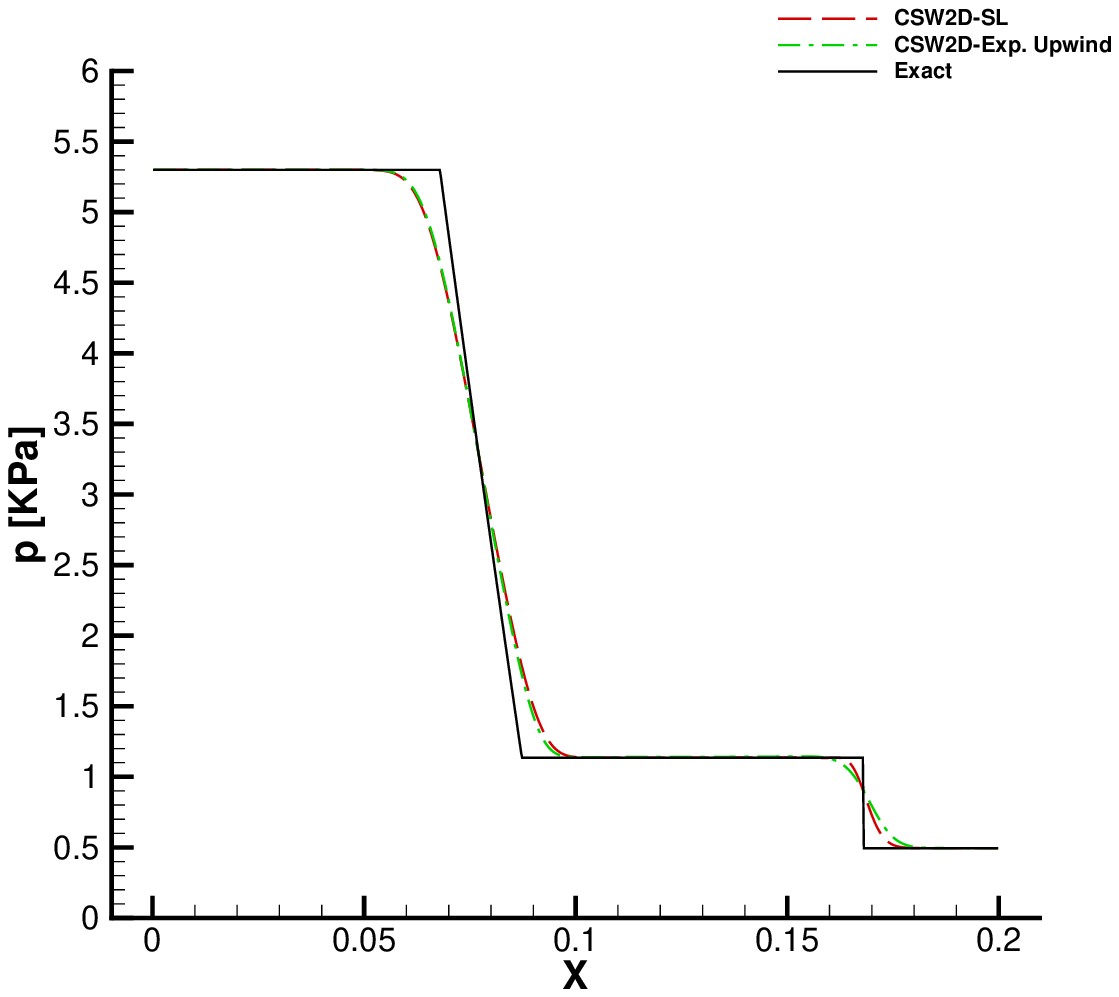}
		\end{tabular} 
		\caption{3D view of the tube and its deformation, mass flow, non-dimensional cross-sectional area and pressure, at the final time $t=0.013$ respectively from top left to bottom right.} 
		\label{fig.RP1alessia}
	\end{center}
\end{figure}
Figure \ref{fig.RP1alessia} shows the mesh  with the deformation of the tube (top left panel),
as well as a few relevant profiles of the solution at the final time, 
with  a direct comparison with \cite{Lucca2023}. 
In our case the data is reported for all the $50$ angles. 
Due to the homogeneity of the tube,  the flow remains axially symmetric all along the simulation, 
but in general this is not required, i.e. $\eta=\eta(r,\theta)$. 
Our algorithm shows an excellent agreement with respect to the exact solution of the problem. 
%=================================================================
\section{Conclusions}
\label{sec.conclusions}
We have presented a new covariant formulation of the shallow water equations, see Eqs.~\eqref{eq:CSW-1-f}-\eqref{eq:CSW-2-f}, which minimizes the formal corrections
with respect to the classical form in Cartesian coordinates. In particular, no Christoffel symbol enters the equations.
The new formulation is quite general, allowing for applications to arbitrary three-dimensional manifolds, 
although in this paper we have limited our attention
to two-dimensional manifolds covered by orthogonal coordinates, hence with a diagonal metric.
The equations have been discretized by means of the semi-implicit schemes developed over the years by Casulli and collaborators (see \cite{Casulli2022} for a review). The computational approach is particularly efficient and it combines the following key features
\begin{itemize}
	\item An implicit discretization for the gradient of surface
	elevation in the momentum equations and for the velocity in the continuity equation.
	\item The capability of treating coordinate singularities (such as those at the earth poles) by means of the same Newton approach 
	that accounts for wetting and drying in flat geometries.
 	\item A natural well-balancing property to preserve stationary solutions up to machine precision.
	\item The possibility of activating a sub-grid discretization for a higher accurate representation of the bathymetry.
\end{itemize}
The new approach has been verified over a number of stringent tests for the classical shallow water equations, including: the propagation of a smooth wave over waterland across the poles, the well balanced property of the entire earth oceans, 
standard Riemann problems, the simulation of a steady state geostrophic flow and the simulation of blood flow in an artery with deformation. In particular, the method can handle coordinate singularities (such as those at the poles in spherical coordinates)
with no need to introduce any special treatment, but simply due to a few built-in properties that are inherent to the numerical scheme: a semi-Lagrangian discretization of the advective terms, and precise mass conservation in  a wetting and drying framework.

Because of its flexibility, several more realistic applications become possible, such as oceanographic simulations of the entire planet, as well as the modeling of the full artery system.

%=================================================================
\section{Acknowledgments}
We are deeply indebted to Prof. Vincenzo Casulli, whose ideas and intuitions have inspired our investigation.
M.T. gratefully acknowledges the support received from the European Union with the ERC Starting Grant \textit{ALcHyMiA} (grant agreement No. 101114995).
Views and opinions expressed are however those of the author only and do not necessarily reflect those of the European Union or the European Research Council Executive Agency. 
Neither the European Union nor the granting authority can be held responsible for them.
M.T. is also member of the INdAM GNCS group in Italy;
%----------------------------------------------------

%=========================================================================
\appendix

\section{Derivation of Eqs.~\eqref{eq:CSW-1}-\eqref{eq:CSW-2}. }
\label{sec:appendixA}
The continuity equation  \eqref{eq:cov-1} is first integrated along the vertical direction, dividing by the constant term $\rho$. We immediately get:
\begin{equation}
	\label{eq:cov-1bis}
\frac{\partial H}{\partial t}+ \nabla_\beta m^\beta=0\,.
\end{equation}
Now we use the standard expression for the covariant divergence of a generic vector $m^\alpha$ \citep{Schutz80}
\begin{equation}
	\nabla_\beta m^\beta=\frac{1}{\sqrt{\gamma}}\partial_\beta(\sqrt{\gamma}m^\beta)\,,
\end{equation}
which allows to obtain Eq.~\eqref{eq:CSW-1}.
We now focus on the momentum equation \eqref{eq:cov-2}. Due to the incompressibility assumption, we can move $u^\beta$ into the covariant derivative, to obtain 
\begin{equation}
	\frac{\partial u^\alpha}{\partial t}+   \nabla_\beta(u^\beta u^\alpha)+ \nabla_\beta (p \gamma^{\alpha\beta})=0\,.
\end{equation}
We then average along the vertical direction (i.e. normally to the manifold), performing the same approximations of local velocities with their vertically averages as in \cite{Casulli2022}, recalling that $v^\alpha=\frac{1}{H}\int_{b}^\eta u^\alpha\,dz$ 
\begin{align}
	&\Longrightarrow \text{(vertical averaging)}\,\,\,\,	\frac{\partial (H v^\alpha)}{\partial t}+   \nabla_\beta(H v^\beta v^\alpha)+ \gamma^{\alpha\beta} \int_b^\eta \nabla_\beta [g(\eta-z)] \,dz=0\,,\\
	&\Longrightarrow \text{(expand integration)}\,\,\,\,	\frac{\partial (H v^\alpha)}{\partial t}+   \nabla_\beta(H v^\beta v^\alpha)+ \gamma^{\alpha\beta}\left[\nabla_\beta\int_b^\eta  g(\eta-z) \,dz +g(\eta-b)\partial_\beta b \right]\\
	&\Longrightarrow \text{(use $H=\eta-b$)}\,\,\,\,	\frac{\partial (H v^\alpha)}{\partial t}  +   \nabla_\beta(H v^\beta v^\alpha)  + \nabla_\beta \left(\frac{1}{2}g H^2  \gamma^{\alpha\beta}\right) + g H \gamma^{\alpha\beta}\partial_\beta b=0\,,\\
	\label{eq:tmp}
	&\Longrightarrow \text{(use $m^\alpha=H v^\alpha$)}\,\,\,\,	\frac{\partial m^\alpha}{\partial t}  +   \nabla_\beta\left(\frac{m^\beta m^\alpha}{H}   + \frac{1}{2}g H^2 \gamma^{\alpha\beta}\right) + g H \gamma^{\alpha\beta}\partial_\beta b=0\,.
\end{align}	
The above equation is the same as Eq.~(1b) by \cite{Carlino2023}.
At this point we introduce the auxiliary symmetric tensor $\tilde T^{\alpha\beta}=\frac{ m^\alpha m^\beta}{H}   +\frac{1}{2}g H^2 \gamma^{\alpha\beta}$ and we recall the fundamental identity for the covariant divergence of any symmetric tensor, i.e. [see Sect. 2.7.5 in \cite{Carmeli2001}]
\begin{equation}
	\label{eq:divsym}
	\nabla_\beta \tilde T^\beta_\alpha=\frac{1}{\sqrt{\gamma}}\partial_\beta(\sqrt{\gamma}\, \tilde T^\beta_\alpha )-\frac{1}{2}\tilde T^{\mu\nu}\partial_\alpha \gamma_{\mu\nu}\,.
\end{equation}
We also recall the fundamental identity which follows from $\nabla_\alpha\gamma_{\mu\nu}=0$, namely
\begin{equation}
	\label{eq:diff}
	\gamma^{\mu\nu}\partial_\alpha \gamma_{\mu\nu}=2\frac{\partial_\alpha \sqrt\gamma}{\sqrt\gamma}\,.
\end{equation}
Hence Eq.~\eqref{eq:tmp}, written with covariant indices, provides 
\begin{align}
	&	\frac{\partial m_\alpha}{\partial t}  +   \nabla_\beta\left(\frac{m^\beta m_\alpha}{H}   + \frac{1}{2}g H^2  \delta_\alpha^{\beta}\right) + g H \partial_\alpha b=0\,,\\
	&\Longrightarrow \text{(use Eq.~\eqref{eq:divsym})}\,\,\,\,	\frac{\partial m_\alpha}{\partial t}  +  \frac{1}{\sqrt{\gamma}}\partial_\beta \left(\sqrt{\gamma}\frac{m^\beta m_\alpha}{H}   + \sqrt{\gamma} \frac{1}{2}g H^2 \delta_\alpha^{\beta}\right)-\frac{1}{2}\left(\frac{m^\mu m^\nu}{H}   + \frac{1}{2}g H^2 \gamma^{\mu\nu}\right)\partial_\alpha\gamma_{\mu\nu} + g H \partial_\alpha b=0\,,\\
	&\Longrightarrow \text{(multiply by $\sqrt\gamma$)}\,\,\,\,	\frac{\partial( \sqrt\gamma m_\alpha)}{\partial t}  +  \partial_\beta \left(\sqrt{\gamma}\frac{m^\beta m_\alpha}{H}\right)   + \frac{1}{2}\partial_\alpha (g\sqrt{\gamma} H^2) -\frac{1}{2}\sqrt\gamma\frac{m^\mu m^\nu}{H}\partial_\alpha\gamma_{\mu\nu}  \nonumber \\ 
	& \hspace{3.3cm} - \frac{1}{4}\sqrt\gamma g H^2 \gamma^{\mu\nu}\partial_\alpha\gamma_{\mu\nu} + \sqrt\gamma g H \partial_\alpha b=0\,,\\
	&\Longrightarrow \text{(use Eq.~\eqref{eq:diff})}\,\,\,\,\frac{\partial( \sqrt\gamma m_\alpha)}{\partial t}  +  \partial_\beta \left(\sqrt{\gamma}\frac{m^\beta m_\alpha}{H}\right)   + \frac{1}{2}g\partial_\alpha (\sqrt{\gamma} H^2)  +\frac{1}{2}\sqrt{\gamma}H^2\partial_\alpha g 
	-\frac{1}{2}\sqrt\gamma\frac{m^\mu m^\nu}{H}\partial_\alpha\gamma_{\mu\nu}  \nonumber \\ 
	& \hspace{3.3cm} - \frac{1}{2}g H^2 \partial_\alpha\sqrt\gamma +  \sqrt\gamma g H \partial_\alpha b=0\,,\\	
	&\Longrightarrow \text{(clean terms)}\,\,\,\,\frac{\partial( \sqrt\gamma m_\alpha)}{\partial t}  +  \partial_\beta \left(\sqrt{\gamma}\frac{m^\beta m_\alpha}{H}\right)   + g\sqrt\gamma H \partial_\alpha  \eta =\frac{1}{2}\sqrt\gamma\left[\frac{m^\mu m^\nu}{H}\partial_\alpha\gamma_{\mu\nu}
	-H^2\partial_\alpha g
	\right]\,,
	\label{eq:last}
\end{align}
which coincides with Eq.~\eqref{eq:CSW-2} in the text, except for the last term on the right hand side of Eq.~\eqref{eq:last} expressing the gravity gradient. Such a term, which has been intentionally omitted in Eq.~\eqref{eq:CSW-2}, is very important for simulating tidal waves and its effects will be considered in a dedicated work.
%=============================================================================
%==========  B I B L I O \gamma R A P H Y

%=============================================================================
% Bibliography with names of authors in text
\setcitestyle{numbers}
\bibliographystyle{elsarticle-harv-nourl}
\bibliography{./references.bib}
%=============================================================================
% Bibliography with references addressed through numbers 
%\bibliographystyle{plain}
%\bibliography{./references.bib}
%=============================================================================

% alternative to remove hyperlinks in references
%\printbibliography

\end{document}